\documentclass[aps,twocolumn,prc,superscriptaddress,showpacs]{revtex4}%
\usepackage{graphicx}
\usepackage{dcolumn}
\usepackage{bm}
\usepackage{color}
\usepackage{amsmath}
\usepackage{amsfonts}
\usepackage{amssymb}
\usepackage{wasysym}%
\providecommand{\U}[1]{\protect\rule{.1in}{.1in}}
\begin{document}
\title{Relativistic mean field interaction with density dependent meson-nucleon
vertices based on microscopical calculations}
\author{X. Roca-Maza\thanks{\texttt{e-mail: roca@ecm.ub.es}}}
\affiliation{Departament d'Estructura i Constituents de la Mat\`eria and Institut de Ci\`encies del Cosmos, Facultat de F\'isica,
Universitat de Barcelona, Diagonal 647, E-08028 Barcelona, Spain}
\affiliation{INFN, sezione di Milano, via Celoria 16, I-20133 Milano, Italy}
\author{X. Vi\~nas \thanks{\texttt{e-mail: xavier@ecm.ub.es }}}
\affiliation{Departament d'Estructura i Constituents de la Mat\`eria and Institut de Ci\`encies del Cosmos, Facultat de F\'isica,
Universitat de Barcelona, Diagonal 647, E-08028 Barcelona, Spain}
\author{M. Centelles \thanks{\texttt{e-mail: mario@ecm.ub.es}}}
\affiliation{Departament d'Estructura i Constituents de la Mat\`eria and Institut de Ci\`encies del Cosmos, Facultat de F\'isica,
Universitat de Barcelona, Diagonal 647, E-08028 Barcelona, Spain}
\author{P. Ring \thanks{\texttt{e-mail: ring@ph.tum.de}}}
\affiliation{Physikdepartment, Technische Universit\"at M\"unchen, D-85748 Garching,
Germany }
\author{P. Schuck \thanks{\texttt{e-mail: schuck@ipno.in2p3.fr }}}
\affiliation{Institut de Physique Nucl\'eaire, CNRS, UMR8608, Orsay, F-91406, France}%
\affiliation{Universit\'e Paris-Sud, Orsay, F-91505, France}%
\affiliation{Laboratoire de Physique et Mod\'{e}lisation des Milieux
Condens\'{e}s, Grenoble, F-38042, France}

\date{\today}

\begin{abstract}
Although ab-initio calculations of relativistic Brueckner theory lead to large
scalar isovector fields in nuclear matter, at present, successful versions of
covariant density functional theory neglect the interactions in this channel.
A new high precision density functional DD-ME$\delta$ is presented which includes
four mesons $\sigma$, $\omega$, $\delta$, and $\rho$ with density dependent
meson-nucleon couplings. It is based to a large extent on microscopic ab-initio
calculations in nuclear matter. Only four of its parameters are determined
by adjusting to binding energies and charge radii of finite nuclei. The other
parameters, in particular the density dependence of the meson-nucleon vertices,
are adjusted to non-relativistic and relativistic Brueckner calculations of
symmetric and asymmetric nuclear matter. The isovector effective mass
$m_{p}^{\ast}-m_{n}^{\ast}$ derived from relativistic Brueckner theory is used to
determine the coupling strength of the $\delta$-meson and its density dependence.

\end{abstract}

\maketitle

\bigskip


\section{Introduction}


Structure properties of nuclei described in the framework of
effective mean-field interactions are remarkably successful over
almost the entire periodic table~\cite{VB.72,DG.80,NL1,NL3,FST.97a,TW.99,
DD-ME1,GSB.03,TP.05,DD-ME2,NL3*,BSV.08}. Relativistic and
non-relativistic versions of this approach enable an effective
description of the nuclear many-body problem as an energy density
functional. These energy functionals are usually adjusted to a
variety of finite nuclei and infinite nuclear matter properties.
Although all these effective interactions are based on the mean-field
approach, some differences will generally appear between them due to
the specific ansatz of the density dependence adopted for each
interaction. For instance, predictions in the isovector channel
of existing functionals differ widely from one another and, as a
consequence, the density dependence of the symmetry energy is far from
being fully determined. This has an impact on finite nuclei properties as,
for example, the neutron skin thickness. Mean field models which accurately
describe the charge radius in $^{208}$Pb, predict neutron radii between 6.6 and 5.8 fm.
For these reasons, one of the main goals in Nuclear Physics is to build
a universal density functional theory based on microscopical calculations
~\cite{LNP.641,DFP.10}. This functional should be able to explain
as many as possible measured data within the same parameter set and
to provide reliable predictions for properties of nuclei far from
stability not yet or never accessible to experiments in the laboratory.
It should be derived in a fully microscopic way from the interactions
between bare nucleons. At present, however, attempts to derive such a
density functional provide only qualitative results for two reasons:
first the three-body term of the bare interaction is not known well enough
and second the methods to derive such a functional are not precise enough
to achieve the required accuracy. Note that a 1 \permil\ error in the binding energy
per particle of symmetric nuclear matter leads to an error of several MeV
in the binding energy of heavy nuclei, an error which is an order of magnitude
larger than required by astrophysical applications. Therefore, at present
the most successful functionals are derived either in a fully phenomenological
way from a very large set of experimental data, as for instance more than 2000
nuclear binding energies~\cite{GSB.03} or, more recently, by an adjustment to
a combination of microscopic results and to a set of characteristic
experimental data~\cite{FKV.06,DD-PC1,UNEDF,GCP.10}.


With the same goal, in the more recent past a different approach was
tried. Baldo, Schuck and Vi\~{n}as employed the same route as in
condensed matter physics and constructed a functional
(BCP)~\cite{BSV.08} where the bulk part is based on the fully
microscopic calculations of Ref.~\cite{BMS.04}. In this more
fundamental calculations, Baldo and collaborators investigate the
nuclear and infinite neutron matter on the basis of the
Bethe-Brueckner approach including three-body correlations of the
Bethe-Faddeev type~\cite{Ba99}.
Their results for the Equation of State (EoS) of symmetric and neutron
matter are believed to be among
the most accurate in the literature. Then, the BCP functional took
as a benchmark the calculations of Ref.~\cite{BMS.04} by means
of a polynomial fit. In this way an accurate and analytic EoS as a
function of neutron and proton densities were constructed covering
the whole range from symmetric nuclear matter to pure neutron matter
in density ranges from zero to about two times saturation density.
Subsequently a finite range surface term dependent on three
parameters together with the strength of the spin-orbit term (four
parameters in total) were added to the bulk part of the functional.
The pairing correlations needed to describe open-shell nuclei were
accounted by a density dependent $\delta$-force with an effective
mass equal to the nucleon mass that simulates pairing calculations in
symmetric nuclear matter computed with the Gogny
interaction~\cite{Ga99} in the $T=1$ channel. Adjusting these free
parameters to some selected nuclear experimental data yielded
excellent results for nuclear masses and reasonable charge radii of
the whole nuclear chart. In addition it has also been shown that the
deformation properties of BCP functionals are similar to the ones
found using the Gogny D1S force~\cite{RBS.08,RBS.10} in spite of the
fact that both models are clearly different. A recent review on the BCP
functionals can be found in Ref.~\cite{BRS.10}

In general, symmetries of nature help to reduce the
number of parameters and to simplify the description. One of
the underlying symmetries of QCD is Lorentz invariance and therefore
covariant versions of density functionals are of particular interest
in nuclear physics. This symmetry allows one to describe
the spin-orbit coupling, which has an essential influence
on the underlying shell structure in finite nuclei, in a consistent
way. Moreover it also puts stringent restrictions on the number of parameters
in the corresponding functionals without reducing the quality of the
agreement with experimental data. Self-consistent mean-field calculations
starting from relativistic Lagrangians have been very successful in
describing nuclear properties~\cite{Rei.89,Ring.96,SW.97,VALR.05,MTZ.06}.
They arise from a microscopic treatment of the nuclear many-body problem
in terms of nucleons and mesons carrying the effective interaction
between nucleons. Moreover, since the theory is relativistically
invariant and the field and nucleon equations of motion are solved
self-consistently, they preserve causality and provide a
self-consistent description of the spin-orbit term of the nuclear
effective force and of the bulk and surface parts of the interaction.
In addition these functionals include {\it nuclear magnetism}~\cite{KR.89},
i.e.  a consistent description of currents and time-odd fields, important for
odd-mass nuclei~\cite{AA.10}, excitations with unsaturated spins,
magnetic moments~\cite{HR.88} and nuclear rotations~\cite{ARK.00}.
No new parameters are required for the time-odd parts of the mean
fields. In non-relativistic
functionals the corresponding time-odd parts are usually difficult
to adjust to experimental data and even if there are additional
constraints derived from Galilean invariance and gauge symmetry~\cite{DD.95}
these constraints are usually not taken into account in the successful functionals
commonly used in the literature. The earlier versions of covariant density functional
theory were based on the Walecka model~\cite{DT.56,Due.56,Wal.74,SW.86} with
phenomenological non-linear meson-interactions proposed by Boguta and Bodmer~\cite{BB.77}
introducing in this way a phenomenological density dependence~\cite{NL3,TP.05,PK1}.
Later the non-linear models have been replaced by an explicit density
dependence of the meson-nucleon vertices. This density dependence
has first been determined in a phenomenological way~\cite{TW.99,DD-ME1,DD-ME2}
These models have shown considerable improvements with respect
to previous relativistic mean-field models in the description of
asymmetric nuclear matter, neutron matter and nuclei far from the
stability valley. On the other hand one has tried to derive this density
dependence in a microscopic way from Brueckner calculations in nuclear
matter at various densities~\cite{BT.92,HKL.01,SOA.05,HSR.07}.
An example is Density Dependent Relativistic Hadron Field
theory~\cite{HKL.01} where the specific density dependence of
the meson-nucleon vertices is mapped from Dirac-Brueckner calculations
where the in-medium interaction is obtained from nucleon-nucleon
potentials consistent with scattering experiments. Therefore,
if this ansatz is adopted, the effective theory is derived fully
from first-principle calculations. Of course, the accuracy of
the results obtained in this way is by no means satisfactory for
modern nuclear structure calculations and a fit of additional
free parameters is still needed. This fact allows to constrain
the different possibilities and keeps the compatibility, at
least theoretically, with more fundamental calculations of
infinite nuclear matter.

As mentioned, there exist \textsl{ab initio} calculations of the
nuclear EoS over a wide range of nucleon densities, \textsl{i.e.} far
from densities currently reachable at the laboratory. In this sense,
apart from the experimental data needed in the fitting procedure for
determining an effective interaction, further steps on building a
universal density functional may need to implement such \textsl{ab
initio} information as it was done in the BCP model. Therefore, this
EoS calculated from first principles can be understood as a temporary
benchmark at supra- and sub-saturation densities of the energy per
particle at different asymmetries. Furthermore, from a theoretical
point of view, consistency is desirable between predictions of both
theories. Regrettably, to make them compatible is not only a problem
of including the EoS derived from realistic nucleon-nucleon
potentials within the fitting procedure. It is also a problem of
taking into account the proper density dependence in the different
terms of the functional. For that, first-principle calculations are
also thought to be the best candidate to help in building a universal
energy density functional, at least the bulk part of it since
many-nucleon calculations are not feasible yet. Hence, for an
effective and self-consistent treatment of the nuclear many body
problem, we propose here a novel and improved relativistic
mean-field interaction with an explicit density dependence of the
meson-nucleon vertices in all the four spin-isospin channels
compatible with fully microscopical calculations.

The essential breakthrough of density functional methods in the
description of quantum mechanical many-body problems was Kohn-Sham
theory~\cite{KS.65,KS.65a}, where the exact density functional $E[\rho]$
of Hohenberg and Kohn~\cite{HK.64} was mapped in an exact way on
an effective potential $V_{\rm eff}$ in a single particle Schroedinger
equation, which forms the starting point of all modern applications
of density functional theory. In covariant density functional theory
this effective potential corresponds to the self-energy in the Dirac
equation, which can be decomposed into four channels characterized
by the relativistic quantum numbers of spin and isospin, the scalar
isoscalar channel ($J=0,T=0$), the vector isoscalar channel ($J=1,T=0$)
the scalar isovector channel ($J=0,T=1$) and the vector isovector
channel ($J=1,T=1$). In the Walecka model these channels are
connected with the exchange of mesons carrying the corresponding quantum
numbers. 

However, in nearly all the present successful phenomenological applications
of covariant density functional theory to nuclear structure based on the relativistic 
Hartree model only the three mesons $\sigma$ , $\omega$ , and $\rho$ are taken into
account. The scalar isovector $\delta$-meson ($a_0 (980)$) causing a splitting of the
effective mass between protons and neutrons is neglected, because it has
turned out that usual data such as binding energies and radii of finite nuclei
do not allow to distinguish scalar and vector fields in the isovector channels. 
Allowing independent parameters for the $\rho$- and the $\delta$-mesons leads
to redundancies in the fit. By the same reasons also modern relativistic Hartree-Fock 
\cite{long2006,long2007,long2008} and Hartree-Fock-Bogoliubov \cite{long2010} calculations neglect
the $\delta$-meson in the Lagrangian and, as a consequence, in the direct term.
Of course the Fock term of these calculations contains also contributions to
the scalar isovector channel. Microscopic investigations
by Huber~\textsl{et al.} ~\cite{Hu96,Hu98} and
phenomenological studies~\cite{KK.97,LG.01,LGB.02,GCT.03,ABM.04,GTT.04,BCG.05,LTG.08,RSR.09,RPP.09,Ala.10}
in the the literature stressed that mean-field models which neglect
the $\delta$-meson are likely to miss important ingredients in describing
properly very asymmetric nuclear matter, in particular at high densities.
The proton fraction of $\beta$-stable matter in neutron stars can increase
and the splitting of the effective mass can affect transport properties in neutron stars
and heavy ion reactions. However, as long as the parameters of this
meson are not fixed, such investigations are somewhat academic.
Therefore we derive in this manuscript the $\delta$-nucleon vertex
and its density dependence from modern microscopic calculations
based on the bare nucleon-nucleon force of the T\"ubingen group~\cite{DFF.07}.

The relativistic mean field model DD-ME$\delta$ obtained in this way
is an extension of the DD-ME model developed by the Munich group~\cite{DD-ME1,DD-ME2}
based on the density dependent relativistic Hartree theory. The DD-ME model has
the following degrees of freedom: the proton, the neutron and three mesons carrying the
nuclear interaction, namely $\sigma$-, $\omega$- and $\rho$-mesons. In addition to
these degrees of freedom, we include here a new one, the $\delta$-meson by the reasons
pointed out before. Since this meson provides a treatment of the isospin more close to
the microscopic investigations we can hope that it improves the reliability of the models
for predictions in nuclei far from stability with large isospin
-– planned to be studied experimentally at the new Rare Ion Beam Facilities~\cite{RBFa}.
Apart from the inclusion of the $\delta$-meson in DD-ME$\delta$ the DD-ME$\delta$
model differs from the earlier DD-ME models in that the parameters of DD-ME
were all adjusted to experimental data based on finite nuclei properties,
whereas those of DD-ME$\delta$ are largely based on microscopic ab-initio calculations
in nuclear matter. Only four of the parameters of DD-ME$\delta$ are fitted to finite nuclei.

The paper is organized as follows: after establishing the
DD-ME$\delta$ model in Sect.~\ref{model}, we discuss in 
Sect.~\ref{fit} the strategies to determine the parameters of the
Lagrangian and compare in Sect.~\ref{results} the results of this novel
effective interaction with the experiment and with the non-relativistic
BCP model~\cite{BSV.08} and the completely phenomenological DD-ME2
model~\cite{DD-ME2}, in particular binding energies, charge radii and
neutron skins in spherical nuclei.


\section{Density Dependent Hadron Field Theory}

\label{model}

\subsection{Lagrangian and equations of motion}

Density dependent relativistic hadron field theory which forms the
basis of the DD-ME$\delta$ interaction has been formulated and
extensively discussed in Refs.~\cite{Le95,Fu95,HKL.01}. Here we
present only the essential features of the mean-field equations of motion
derived from such a theory. The relativistic Lagrangian includes neutrons
and protons represented by the Dirac spinors $\psi$ of the nucleon,
the four mesons ($\sigma$, $\omega$, $\delta$ and $\rho$) carrying the effective
nuclear strong interaction represented by the fields $\sigma$,
$\omega^{\mu}$, $\vec{\delta}$, and $\vec{\rho}^{\mu}$, and the
photon field $A_{{}}^{\mu}$ accounting for the electromagnetic
interaction. The index $\mu$ indicates the time- and space-like
components of the vector fields and the arrow indicates the
vector nature of a field in isospin space. As mentioned, the
$\delta$-meson should be included if one wants to follow the
theoretical indications of Dirac-Brueckner calculations in asymmetric
nuclear matter and so we do. The Lagrangian has the following parts%
\begin{equation}
{\mathcal{L}}={\mathcal{L}}_{_{N}}+{\mathcal{L}}_{M}+{\mathcal{L}}_{int}
\label{L}%
\end{equation}
where ${\mathcal{L}}_{_{N}}$ is the nucleonic free Lagrangian%
\begin{equation}
{\mathcal{L}}_{N}=\bar{\psi}\left(  i\gamma_{\mu}\partial^{\mu}-m\right)
\psi\label{LN}%
\end{equation}
${\mathcal{L}}_{M}$ is the Lagrangian of free mesons
\begin{align}
{\mathcal{L}}_{M}  &  =\frac{1}{2}\left(  \partial_{\mu}\sigma\partial^{\mu
}\sigma-m_{\sigma}^{2}\sigma^{2}\right)  +\frac{1}{2}\left(  \partial_{\mu
}\vec{\delta}\partial^{\mu}\vec{\delta}-m_{\sigma}^{2}\vec{\delta}^{2}\right)
\\
&
-\frac{1}{4}\Omega_{\mu\nu}\Omega^{\mu\nu}-\frac{1}{2}m_\omega^{2}\omega_{\mu
}\omega^{\mu}-\frac{1}{4}\vec{R}_{\mu\nu}\vec{R}^{\mu\nu}-\frac{1}{2}m_\rho%
^{2}\vec{\rho}_{\mu}\vec{\rho}^{\mu}\nonumber\\
&  -\frac{1}{4}F_{\mu\nu}F^{\mu\nu}\nonumber
\end{align}
and ${\mathcal{L}}_{int}$ is the Lagrangian describing the
interactions. Its algebraic expression is
\begin{align}
{\mathcal{L}}_{int}  &  =g_{\sigma}\bar{\psi}\sigma\psi+g_{\delta}\bar{\psi
}\vec{\tau}\vec{\delta}\psi\nonumber\\
&  -g_{\omega}\bar{\psi}\gamma_{\mu}\omega^{\mu}\psi-g_{\rho}\bar{\psi}%
\gamma_{\mu}\vec{\tau}\vec{\rho}^{\mu}\psi-e\bar{\psi}\gamma_{\mu}A^{\mu}%
\psi\nonumber
\end{align}
where $m$ is the nucleon mass (commonly taken as $939$ MeV), the field
strength tensors for the vector fields are
\begin{equation}
\Omega^{\mu\nu}=\partial^{\mu}\omega^{\nu}-\partial^{\nu}\omega^{\mu},
\label{F}%
\end{equation}
and correspondingly $\vec{R}^{\mu\nu}$ and $F^{\mu\nu}$. The
electric charge is $e$ for protons and zero for neutrons. The meson-nucleon
vertices are denoted by ${g}_{i}$ for $i=\sigma$, $\omega$, $\delta$,
and $\rho$.
Since covariance is required and the quantity $\sqrt{j_\mu j^\mu}$ is in
the rest frame identical to the baryon density $\rho = \rho_n + \rho_p$,
the nucleon-meson vertices generally depend on this quantity. Because of
the relatively small velocities the difference between $\sqrt{j_\mu j^\mu}$
and $\rho$ is negligible in all practical applications.
The subindex $n$ or $p$ is used to indicate whether we are considering neutrons
or protons respectively.

The equations of motion are derived from the classical variational principle
and we obtain for the nucleon spinors the Dirac equation%
\begin{equation}
\left[  \gamma_{\mu}(i\partial^{\mu}-\Sigma^{\mu})-m^{\ast}\right]  \psi=0
\label{dirac}%
\end{equation}
where $m^{\ast}\equiv m-\Sigma^{s}$ is the effective Dirac nucleon mass and
$\Sigma^{\mu}$ and $\Sigma^{s}$ are the vector and scalar self-energies
defined as follows,
\begin{align}
\Sigma^{s}  &  \equiv{g}_{\sigma}(\rho)\sigma+{g}_{\delta}(\rho)\vec{\tau}%
\vec{\delta}\label{ses}\\
\Sigma^{\mu}  &  \equiv{\Sigma}^{(0)\mu}+\delta_{\mu0}{\Sigma}^{(r)}
\label{sev}%
\end{align}
here, $(0)$ indicates the usual definition of the vector self-energy and $(r)$
the rearrangement term of the vector self-energy
\begin{align}
{\Sigma}^{(0)\mu}  &  \equiv{g}_{\omega}(\rho){\omega}^{\mu}+{g}_{\rho}%
(\rho)\tau^{}_3\rho_3^\mu{+}eA^{\mu}\label{sev0}\\
\Sigma^{(r)}  &  \equiv-\frac{dg_{\sigma}}{d\rho}\sigma\rho^s%
+\frac{dg_{\omega}}{d\rho}\omega^{0}\rho-\frac{dg_{\delta}}{d\rho}\delta^{}_3
\rho^s_3+\frac{dg_{\rho}}{d\rho}\rho^0_3\rho^{}_{3}.
\label{E8}%
\end{align}
Here $eA^0$ is the direct term of the Coulomb potential. As in most RMF models
we neglect in these investigations the Coulomb exchange term which plays
an important role in $pn$-RPA calculations \cite{lian2009}.
The static mean field approximation used throughout this investigation preserves the
third component of the isospin. As a consequence the other two components of the
densities and fields carrying isospin vanish. In Eq.~\ref{E8} and the following
equations $\rho_3^0$ represents the time-like component of the $\rho$-meson field, whereas
$\rho_3=\rho_n-\rho_p$ and $\rho^s_3=\rho^s_n-\rho^s_p$ represent the isovector part
of the baryon density and of the scalar density. The rearrangement term is a contribution to the vector self-energy due
to the density dependence of the meson-nucleon vertices. The equations of motion for the mesons are,
\begin{align}
\left(  \partial_{\nu}\partial^{\nu}+m_{\sigma}^{2}\right)  \sigma &
=-g_{\sigma}(\rho)~\rho^{s}\label{s}\\
\left(  \partial_{\nu}\partial^{\nu}+m_{\omega}^{2}\right)  \omega_{}^{\mu}  &
=+g_{\omega}(\rho)~j_{}^{\mu}\label{o}\\
\left(  \partial_{\nu}\partial^{\nu}+m_{\delta}^{2}\right)  \delta^{}_3 &
=-g_{\delta}(\rho)\rho_{3}^{s}\label{d}\\
\left(  \partial_{\nu}\partial^{\nu}+m_{\rho}^{2}\right)  \rho^{\mu}_3  &
=+g_{\rho}(\rho)~j_{3}^{\mu}\label{r}\\
\partial_{\nu}\partial^{\nu}A_{}^{\mu}  &  =+e~j_{p}^{\mu} \label{f}%
\end{align}
where the different densities and currents are the ground-state expectation
values defined as,
\begin{align}
\rho^{s}\equiv\langle0|\bar{\psi}\psi|0\rangle &  =\rho_{n}^{s}+\rho_{p}%
^{s}\label{ros}\\
j^{\mu}\equiv\langle0|\bar{\psi}\gamma^{\mu}\psi|0\rangle &  =j_{n}^{\mu
}+j_{p}^{\mu}\label{ro}\\
\rho_{3}^{s}\equiv\langle0|\bar{\psi}\tau_{3}\psi|0\rangle &  =\rho_{n}%
^{s}-\rho_{p}^{s}\label{ros3}\\
j_{3}^{\mu}\equiv\langle0|\bar{\psi}\gamma^{\mu}\tau_{3}\psi|0\rangle &
=j_{n}^{\mu}-j_{p}^{\mu}. \label{ro3}%
\end{align}

\bigskip

\subsection{Asymmetric infinite nuclear matter}

\subsubsection{Energy density and pressure}

In infinite nuclear matter we neglect the electromagnetic field.
Because of translational invariance, the Dirac equations can be solved
analytically in momentum space and we obtain the usual plane-wave Dirac
spinors~\cite{Bj65}. Filling up to the Fermi momenta $k_{F\tau}$ for
$\tau=n$ or $p$, we find the densities%
\begin{align}
\rho_{\tau}^{} &=\frac{2}{(2\pi)^{3}}\int_{|k|<k_{F\tau}}d^{3}%
k=\frac{{k_{F\tau}^{3}}}{3\pi^{2}}\label{infro}\\
\rho_{_{\tau}}^{s}&=\frac{2}{(2\pi)^{3}}\int_{|k|<k_{F\tau}}\frac{m_{\tau
}^{\ast}}{E_{\tau}(k)}d^{3}k\nonumber\\
&  =\frac{m_{\tau}^{\ast}}{2\pi^{2}}\left[  k_{F\tau}E_{F\tau}-m_{\tau}%
^{\ast2}{}\ln\left(  \frac{k_{F_{\tau}}+E_{F\tau}}{m_{\tau}^{\ast}}\right)
\right]  \label{infros}%
\end{align}
and the meson fields
\begin{align}
\sigma &  =-\frac{g_{\sigma}(\rho)}{m_{\sigma}^{2}}(\rho_{n}^{s}+\rho_{p}%
^{s})\label{infms}\\
\omega^{0}  &  =+\frac{g_{\omega}(\rho)}{m_{\omega}^{2}}(\rho_{n}^{}+\rho
_{p}^{})\label{infmo}\\
\delta_3 &  =-\frac{g_{\delta}(\rho)}{m_{\delta}^{2}}(\rho_{n}^{s}-\rho_{p}%
^{s})\label{infmd}\\
\rho^{0}_3  &  =+\frac{g_{\rho}(\rho)}{m_{\rho}^{2}}(\rho_{n}^{{}}-\rho_{p}^{{}})
\label{infmr}%
\end{align}
where $E_{\tau}(k)=\sqrt{\mathbf{k}^{2}+m{_{\tau}^{\ast2}}}$ and where the
Fermi energy of neutrons and protons is given by $E_{F\tau}=E_{\tau
}(k_{F\tau})$. Now, we calculate the energy density ($\epsilon$) and pressure ($P$)
from the energy-momentum tensor,
\begin{equation}
T^{\mu\nu}=\sum_{i}\frac{\partial{\mathcal{L}}}{\partial(\partial_\mu\phi
_{i})}\partial^{\nu}\phi_{i}-g^{\mu\nu}{\mathcal{L}}%
\end{equation}
where $\phi_{i}$ runs over all possible fields,
\begin{align}
\epsilon &  =\langle0|T_{}^{00}|0\rangle\\
&  =\frac{1}{4}\left[  3E^{}_{Fn}\rho^{}_{n}+m_{n}^{\ast}\rho_{n}^{s}\right]
+\frac{1}{4}\left[  3E^{}_{Fp}\rho^{}_{p}+m_{p}^{\ast}\rho_{p}^{s}\right]
\nonumber\\
&  +\frac{1}{2}\left[  m_\sigma^{2}\sigma^2+m_\omega^2(\omega^0)^2+m_{\delta}^{2}\delta^{2}_3+m_{\rho}^{2}(\rho_3^0)^2\right]
\nonumber%
\end{align}
and
\begin{align}
\label{pressure}%
P  &  =\frac{1}{3}\sum_{i=1}^{3}\langle0|T_{}^{ii}|0\rangle\\
&  =\frac{1}{4}\left[  E^{}_{Fn}\rho^{}_{n}-m_{n}^{\ast}{\rho_{n}^{s}}\right]
+\frac{1}{4}\left[  E^{}_{Fp}\rho^{}_{p}-m_{p}^{\ast}{\rho_{p}^{s}}\right]
\nonumber\\
&  ~~~~~-\frac{1}{2}\left[  m_{\sigma}^{2}{\sigma}^{2}-
m_{\omega}^{2}(\omega_{}^0)^2+m_{\delta}^{2}\delta^2_3-m_{\rho}^{2}(\rho_{3}^0)^2\right] \nonumber\\
&  ~~~~~~+\left(  \rho_{n}+\rho_{p}\right)  \Sigma^{(r)0}
\nonumber%
\end{align}
Only the pressure has a rearrangement contribution.
We have checked that the pressure derived from the energy-momentum tensor coincides with the
thermodynamical definition: $p=\rho^{2}[\partial(\epsilon/\rho)/\partial\rho]$ and that the
energy-momentum tensor is conserved $\partial_{\mu}T^{\mu\nu}=0$.

\subsubsection{The symmetry energy: $S_2(\rho)$}

Assuming charge symmetry for the strong interaction (the $nn$ and $pp$
interactions are identical but different, in general, from the $np$
interaction), the total energy per particle in asymmetric nuclear matter can
be written as follows,
\begin{equation}
\frac{\epsilon}{\rho}=\frac{E}{A}\equiv e(\rho,\alpha)=e(\rho,\alpha
=0)+S_{2}(\rho)\alpha^{2}+{\mathcal{O}}[\alpha^{4}]%
\label{csymdef1}%
\end{equation}
where $\rho=\rho_n+\rho_p$ is the baryon density and
$\alpha=(\rho_n-\rho_p)/(\rho_n+\rho_p)$ measures the neutron excess.
The term proportional to $\alpha^2$ is the so called symmetry energy
of infinite matter and terms proportional to $\alpha^{4}$ (and
higher) can be neglected to very good approximation. The symmetry
energy $S_2(\rho)$ is defined as,%
\begin{equation}
S_{2}(\rho)=\frac{1}{2}\left(  \frac{\partial^{2}e(\rho,\alpha)}%
{\partial\alpha^{2}}\right)  _{\alpha=0}%
\label{csymdef2}%
\end{equation}
Models including the $\delta$-meson provide a richer description of
the isovector sector of the nuclear strong interaction. For that
reason it is important to understand its effects on asymmetric
nuclear matter and for that we give the analytic expressions for the
symmetry energy of the model discussed in the last
section~\cite{LGB.02,GCT.03}:%
\begin{equation}
S_{2}(\rho)\equiv
S_{2}^{\mathrm{kin}}(\rho)+S_{2}^{\rho}(\rho)+S_{2}^{\delta
}(\rho)\label{csym}%
\end{equation}
with
\begin{align}
S_{2}^{\mathrm{kin}}(\rho) &  =\frac{k_{F}^{2}}{6E_{F}}\\%
S_{2}^{\rho}(\rho) &=\frac{1}{2}\rho\frac{g_{\rho}^{2}}{m_{\rho}^{2}}\\%
S_{2}^{\delta}(\rho)
&=-\frac{1}{2}\rho\frac{g_{\delta}^{2}}{m_{\delta}^{2}}%
\left( \frac{m^{\ast}}{E_{F}}\right)^{2} u_\delta(\rho,m^*)
\label{cspart}%
\end{align}
where for $i=\sigma,\delta$
\begin{equation}
u_i(\rho,m^*) \equiv \frac{1}{1+3\frac{g^2_{i}}{m^2_{i}}\left(
\frac{\rho^{s}}{m^{\ast}}-\frac{\rho}{E_{F}}\right).}\\
\end{equation}
The quantity $u_{\sigma}(\rho,m^*)$ will be needed below.
In these equations we used the fact that for $\alpha=0$ we have
$\rho_{n}=\rho_{p}=\rho/2$,
$\rho_{n}^{s}=\rho_{p}^{s}=\rho^{s}/2$, $\rho_{3}^{s}=0$, $m_{n}^{\ast}%
=m_{p}^{\ast}$ and we have defined $k_{F}^{3}=3\pi^{2}\rho/2$ and $E_{F}%
=\sqrt{k_{F}^{2}+m{^{\ast}}^{2}}$. In symmetric nuclear matter, the effective
mass and the scalar density read
\begin{align}
m^{\ast} &  =m-\frac{g_{\sigma}^{2}}{m_{\sigma}^{2}}\rho^{s}\\
\rho^{s} &  =\frac{m^{\ast}}{\pi^{2}}\left[  k_{F}E_{F}-m{^{\ast}}^{2}%
\ln\left(  \frac{k_{F}+E_{F}}{m^{\ast}}\right)  \right]  ,
\end{align}
respectively. Close to the saturation density $u_\delta \approx 1$ is a very
good approximation and we find in this
case an analytical approximation for $S_{2}^{\delta}(\rho)$%
\begin{equation}
S_{2}^{\delta}(\rho)\approx-\frac{1}{2}\rho\frac{g_{\delta}^{2}}{m_{\delta
}^{2}}\left(  \frac{m^{\ast}}{E_{F}}\right)^{2}%
\label{cdapp}%
\end{equation}
and, therefore, the contribution to the symmetry energy coming from the
nuclear strong interaction (potential part) as described by this kind of
models can be written in the simple form,
\begin{align}
S_{2}^{\text{pot}}(\rho) &  =S_{2}^{\rho}(\rho)+S_{2}^{\delta}(\rho
)\nonumber\\
&  \approx\frac{1}{2}\rho\left[  \frac{g_{\rho}^{2}}{m_{\rho}^{2}}%
-\frac{g_{\delta}^{2}}{m_{\delta}^{2}}\left(  \frac{m^{\ast}}{E_{F}}\right)
^{2}\right]  \label{cpotapp}%
\end{align}
The symmetry energy is often expanded around the saturation density
$\rho^{}_{\rm sat}$%
\begin{equation}
S_{2}(\rho) = J + \frac{L}{3\rho^{}_{\rm sat}}(\rho-\rho^{}_{\rm sat})%
 + \frac{K_{\rm sym}}{18\rho^2_{\rm sat}}(\rho-\rho^{}_{\rm
sat})^2+\dots
\label{esymmexp}%
\end{equation}
where $J$ is the symmetry energy at saturation, and $L$ and $K_{\rm sym}$ are proportional,
respectively, to the slope and the curvature of the symmetry energy at saturation.

Using the analytical expressions (\ref{cspart}) we find
\begin{equation}
L(\rho) \equiv 3\rho\frac{d S_2(\rho)}{d\rho} = L^{\mathrm{kin}}(\rho) +
L^{\rho}(\rho) + L^{\delta}(\rho)%
\label{E39}
\end{equation}
with
\begin{eqnarray}
L^{\mathrm{kin}}(\rho)     &=&  S^{\mathrm{kin}}_2 \left( 2-\frac{k^2_{F}}{E^2_F}-\frac{3m^{*2}}{E^2_F}w \right)\\%
L^{\rho}(\rho)  &=& S_2^{\rho} \left(3 + 6\frac{\rho}{g_\rho}\frac{\partial g_\rho}{\partial \rho}\right) \\%
L^{\delta}(\rho)&=&S_2^{\delta}\left[3 + 6\frac{\rho}{g_{\delta}}\frac{\partial g_\delta}{\partial\rho}
- \frac{2k^2_{F}}{E^2_{F}} + 6\left(1-\frac{m^{*2}}{E^2_{F}}\right)w\right.\nonumber \\%
&-& \left. 3\frac{g^2_{\delta}}{m^2_{\delta}}u_{\delta}
\left(2v\left(\frac{\rho}{g_{\delta}} \frac{\partial
g_{\delta}}{\partial \rho} + w\right)   + \rho\frac{k^2_{F}}{E^3_{F}}\left(1-3w\right)\right) \right]  \nonumber\\
\label{lpart}
\end{eqnarray}
where the functions $u_i$, $v$ and $w$ depend on $\rho$ and $m^*$:
\begin{eqnarray}
v(\rho,m^*) &\equiv& 3\left(\frac{\rho^s}{m^*}-\frac{\rho}{E_{F}}\right)\\
w(\rho,m^*) &\equiv& \frac{\rho}{m^*}\frac{\partial m^*}{\partial \rho}\\
&=& -\frac{g^2_{\sigma}}{m^2_{\sigma}}u_{\sigma}
\left(2\frac{\rho^s}{m^*}\frac{\rho}{g_{\sigma}}\frac{\partial g_{\sigma}}{\partial
\rho} + \frac{\rho}{E_{F}}\right)\nonumber
\end{eqnarray}
The strength of the $\sigma$- and $\omega$-nucleon vertices is quite
well determined by experimental data as compared with the strength of
the isovector meson-nucleon vertices. On the other side, with only the
$\rho$-nucleon vertex, one is able to reproduce properties of finite nuclei~\cite{VNR.03}
and to account for the symmetry energy around saturation in rather good agreement with
available empirical indications. However, to reproduce nucleon-nucleon scattering
measurements in the vacuum, one needs to incorporate a
scalar-isovector meson into the parameterization of the two-body
nuclear interaction~\cite{Mac.89}. Microscopic derivations of the
nuclear fields using relativistic Brueckner
theory~\cite{Hu96,Hu98,HKL.01,DFF.05a,DFF.05b,KBT.06,DFF.07} or
non-relativistic Brueckner theory~\cite{SOA.05,HSR.07} show clearly
that the scalar field in the nuclear interior has an isovector part.
These reasons motivate one to incorporate the $\delta$-meson
also in models of covariant density functional theory and to study
its influence on properties such as the
symmetry energy, the effective mass splitting between protons and
neutrons in asymmetric matter, the isospin dependence of the spin-orbit
potential and the spin-orbit splittings far from stability.

\subsection{Density dependence of the meson-nucleon vertices}

Here we describe the density dependence of the meson-nucleon
vertices used for the new interaction DD-ME$\delta$. We start from
modern fully microscopic calculations in symmetric nuclear matter and
pure neutron matter at various densities and try to determine the
density dependence of the vertices $g_{i}(\rho)$ by fitting to those
data. Of course, it is well known that successful density
functionals can, at present, not be determined completely from
ab-initio calculations. Therefore, we introduce in the fit not
only results of microscopic calculations but also a set of data on
binding energies and radii in specific finite nuclei.

In a first step we have to choose a form of the density dependence of
the various vertices, which is flexible enough to reproduce the
microscopic calculations. In Refs.~\cite{BT.92,JL.98a}
the meson-nucleon vertices of density dependent RMF theory have been
related to the scalar and vector self-energies obtained from Dirac-Brueckner
(DB) calculations in infinite nuclear matter.
The density dependence deduced from DB calculations is
\begin{equation}
g_{i}(\rho)=g_{i}(\rho_{\text{sat}})f_{i}(x)\quad\mathrm{for}\quad
i=\sigma,\omega,\delta,\rho\label{dd}%
\end{equation}
where $\rho_\text{sat}$ is the saturation density of symmetric
nuclear matter and $x=\rho/\rho_{\text{sat}}$. For the functions $f_{i}(x)$ we follow
Refs.~\cite{TW.99,DD-ME1,DD-ME2} and use the Typel-Wolter ansatz:
\begin{equation}
f_{i}(x)=a_{i}\frac{1+b_{i}(x+d_{i})^{2}}{1+c_{i}(x+e_{i})^{2}} .
\label{fx}%
\end{equation}
As in Refs.~\cite{DD-ME1,DD-ME2,DD-PC1} we use the value $\rho_{\text{sat}}=0.152$ fm$^{-3}$.
In fact, this choice is very close to the saturation density obtained in the following fit.
As we see from the ansatz~(\ref{fx})
the actual value of $\rho_\text{sat}$ is irrelevant for the calculations. It can be completely
absorbed in the values of the parameters $b_i$, $c_i$, $d_i$, and $e_i$. It is only used to
make them dimensionless. By definition, the parameters $a_{i}$ are
constrained by the condition $f(1)=1$. In the earlier
applications~\cite{TW.99,DD-ME1,DD-ME2} this ansatz was only used for
the $\sigma$- and $\omega$-meson. The density dependence of the
isovector coupling $g_{\rho}(\rho)$ was described by an exponential
and the $\delta$-meson was neglected. Here we use the same ansatz
(\ref{fx}) also for the isovector mesons $\delta$ and $\rho$. This
turned out to be necessary in order to obtain a density dependence of
the $\delta$-coupling similar to that derived from microscopic
\textit{ab-initio} calculations in Refs.~\cite{JL.98a,Jl.98b,HKL.01}.
We impose as in Ref.~\cite{TW.99} the constraints
$e_{\sigma}=d_{\sigma}$, $e_{\omega}=d_{\omega}$,
$f_{\sigma}^{^{\prime\prime}}(x=1)=f_{\omega}^{^{\prime\prime}}(x=1)$ and
$f_{i}^{^{\prime\prime}}(x=0)=0$. We work with meson masses
$m_{\omega}=783$ MeV, $m_{\delta}=983$ MeV and $m_{\rho}=763$ MeV.
The nucleon mass is $m=939$ MeV. All in all, the model has 14
adjustable parameters. Namely, the 4 coupling constants
$g_{i}(\rho_{\text{sat}})$ in the 4 relativistic channels
(Lorentz-scalar, Lorentz-vector, isoscalar and isovector), 9
parameters describing the density dependence in the functions $f_{i}(x)$,
and the $\sigma$-mass $m_{\sigma}$ allowing for a finite range and a
proper description of the nuclear surface.

\subsection{Calculation of finite nuclei}

The self-consistent results for masses include a
microscopic estimate for the center-of-mass correction:
\begin{equation}
E_{\mathrm{cm}}=-\frac{\langle P^{2}_{\mathrm{cm}}\rangle}{2mA} \label{cm}%
\end{equation}
where $P_{\mathrm{cm}}$ is the total momentum of a nucleus with
$A$ nucleons. The expression
\begin{equation}
r_{\mathrm{c}}=\sqrt{\langle r_{p}^{2}\rangle+ 0.8^{2}}%
\label{rch}%
\end{equation}
is used for the charge radius. The description of open shell nuclei
requires pairing correlations. We introduce this through the BCS approach
with a seniority zero force in the soft pairing window described in Ref.~\cite{BRR.00}. For
the fit of the parameters of the Lagrangian described in the next section
the constant gap approximation~\cite{Vau.73} has been used and the gap
parameters have been derived from the experimental binding energies
by a 3-point formula.

\section{The parameters of the functional DD-ME$\delta$}
\label{fit}

In this section we describe the determination of the parameters of
DD-ME$\delta$. Earlier fits of relativistic Lagrangians have shown
that the usual set of experimental ground state properties in finite nuclei,
such as binding energies and radii do not allow to determine more
than 7 or 8 parameters~\cite{NL3}. Two of them ($g_{\sigma}/m_{\sigma}$ and
$g_{\omega}/m_{\omega}$) determine the saturation energy and the
saturation density of symmetric nuclear matter~\cite{Wal.74}, one of
them ($m_{\sigma}$) is fixed by the radii in finite nuclei and another
one of them ($g_{\rho}/m_{\rho}$) determines the symmetry energy $J$ at
saturation. The additional parameters (as for instance $g_{2}$ and
$g_{3}$ in the nonlinear meson coupling models NL1~\cite{NL1} or
NL3~\cite{NL3} or the three parameters in the ansatz (\ref{fx}) for
density dependence in the isoscalar channel of DD-ME1~\cite{DD-ME1}
and DD-ME2~\cite{DD-ME2}) are determined by the isoscalar surface
properties and are necessary to describe deformations and the nuclear
incompressibility properly. Finally one parameter ($a_{\rho}$ in
DD-ME1 or DD-ME2) is needed to describe the density dependence of the
symmetry energy by a fit to the experimental data on the neutron skin
thickness.
\begin{figure}[h]
\vspace{0.6cm}
\begin{center}
\includegraphics[width=8cm,clip=true]{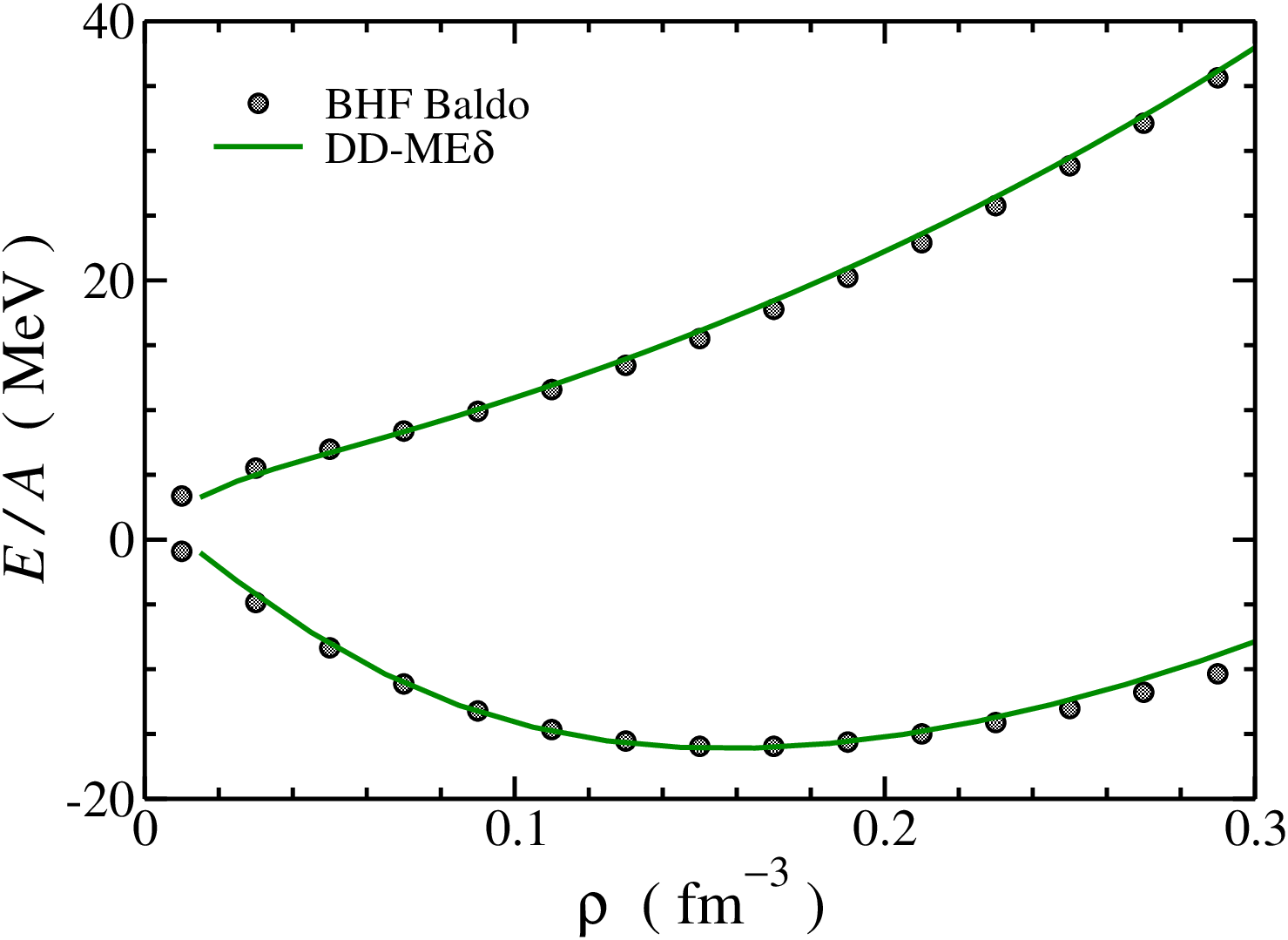}
\end{center}
\vspace{-0.5cm}%
\caption{(Color online) Equation of state for symmetric nuclear matter and for
pure neutron matter as a function of the
nucleon density. The dots represent the
predictions of the BHF calculations~\cite{BMS.04} and the line
our fit to reproduce this data.}%
\label{fig1}%
\end{figure}

\begin{figure}[h]
\vspace{0.6cm}
\begin{center}
\includegraphics[width=8cm,clip=true]{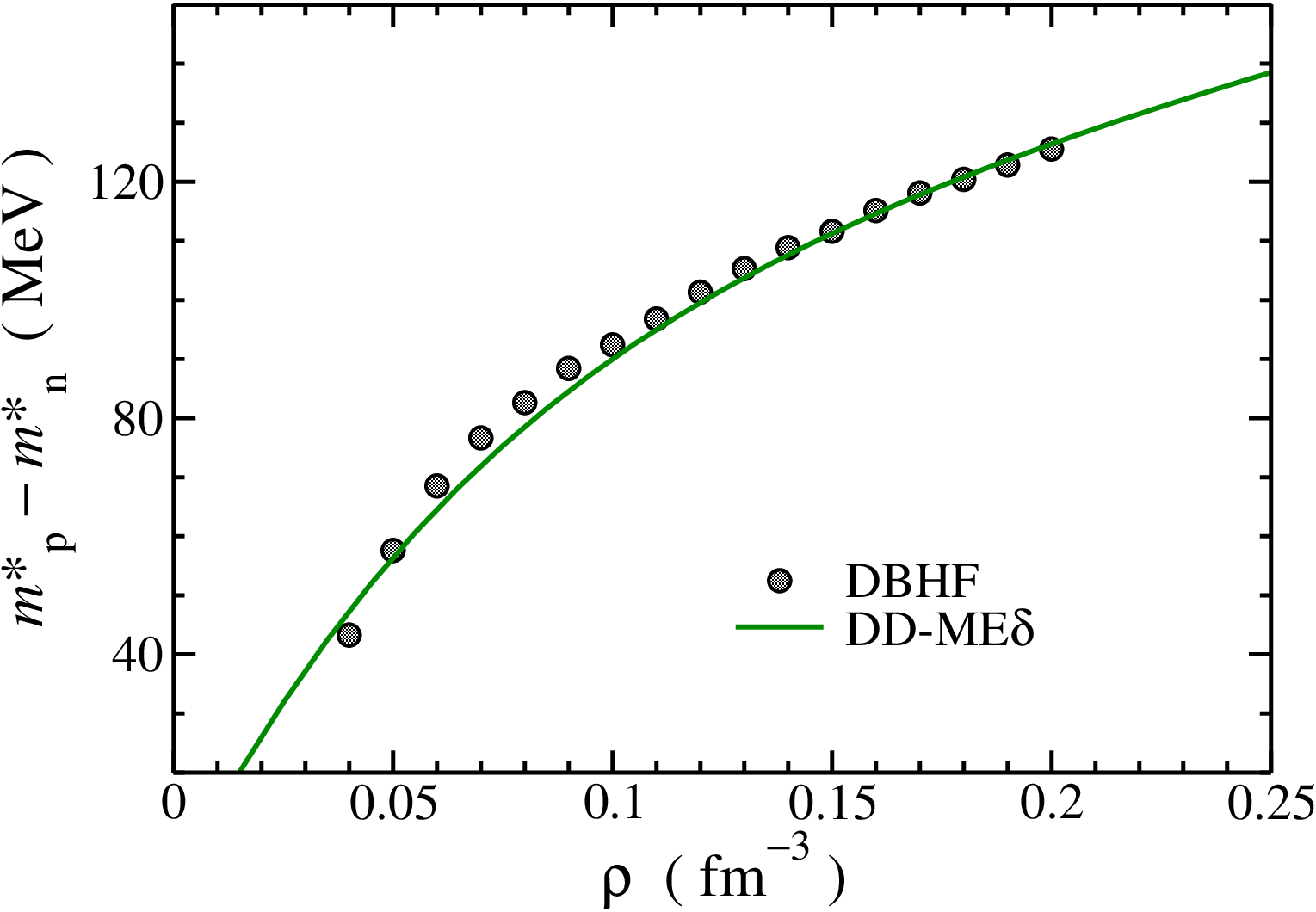}
\end{center}
\vspace{-0.5cm}%
\caption{(Color online) Proton-neutron effective mass splitting as a function of the
nucleon density in pure neutron matter. The dots represent the
predictions of the DBHF calculations~\cite{DFF.07} and the line
our fit to reproduce this data.}%
\label{fig2}%
\end{figure}

In order to calibrate the 14 free parameters of the DD-ME$\delta$ functional
we therefore added pseudo-data in the form of results of modern microscopic
non-relativistic and relativistic Brueckner calculations. To this end,
we selected the EoS of symmetric nuclear matter and of neutron matter
(see Fig.~\ref{fig1}) derived by Baldo et al.~\cite{BMS.04} in a
state-of-the-art non-relativistic Brueckner calculation including
relativistic corrections and three-body forces. We also used as a
benchmark the isovector part of the effective Dirac mass
$m_{p}^{\ast}-m_{n}^{\ast}$ (see Fig.~\ref{fig2}) derived by the
T\"{u}bingen group~\cite{DFF.07} in relativistic Dirac-Brueckner theory.
The use of non-relativistic results for the EoS and of relativistic
results for the isovector effective mass may seem somewhat arbitrary.
However we have to keep in mind that the non-relativistic calculations
of the Catania group are more sophisticated than presently available
Dirac-Brueckner calculations, because they include not only relativistic
effects but also three-body forces. On the other side it is very complicated
to deduce Dirac masses from a non-relativistic calculation which does not
distinguish between Lorentz scalars and vectors. This is in principle possible~\cite{SOA.05,HSR.07}, but it is difficult and connected with additional uncertainties. With this caveat in mind, we decided to use a
reliable relativistic Brueckner calculation~\cite{DFF.07}
providing directly the effective Dirac masses $m_{p}^{\ast}$ and
$m_{n}^{\ast}$ and their difference, a quantity directly connected
with the scalar isovector part of the self energy.
The isovector part of the effective Dirac mass
$m_{p}^{\ast}-m_{n}^{\ast}$ depends only on the $\delta$-meson.
It vanishes for all the conventional Lagrangians without $\delta$-meson.
The density dependence of this quantity is
therefore the optimal tool to get information about the density
dependence of the $\delta$-meson vertex $g_{\delta}(\rho)$.

Keeping this in mind, we determine 10 of the 14 parameters in
the Lagrangian of DD-ME$\delta$ by these pseudo-data obtained from
\textit{ab-initio} calculations of  nuclear matter. These parameters
define the density dependence for the various
meson-nucleon vertices (9 parameters) and the strength
$g_{\delta}(\rho_{\text{sat}})$ of the $\delta$-meson. Only a
reduced set of 4 parameters ($g_{\sigma}(\rho_{\text{sat}})$,
$g_{\omega}(\rho_{\text{sat}})$, $g_{\rho}(\rho_{\text{sat}})$, and
$m_{\sigma}$) are fitted to the masses and charge radii of finite
nuclei.

\subsection{Strategy of the parameter fit}
\label{section3a}

Since the mean field equations of motion have to be solved
self-consistently, we need a good starting parameter set before
fixing the $\delta$-meson coupling to the above mentioned
calculations fully. The density-dependent meson coupling model
DD-ME2~\cite{DD-ME2} provides us with an excellent description of nuclei
all over the periodic table. Though DD-ME2 neglects the $\delta$-meson,
it is based in the isoscalar channel on the same ansatz (\ref{fx}).
Therefore, we used DD-ME2 as a starting point of our investigations.
We proceeded in three steps:

\begin{enumerate}
\item[1)] In the first step we performed an overall fit with all the 14
parameters. For the data we have chosen on one side the three
microscopic curves for the EoS in Figs.~\ref{fig1} and ~\ref{fig2}
and on the other side the same set of data of finite nuclei which has
been used in Ref.~\cite{DD-ME2} for the determination of the parameter
set DD-ME2 (see Table II of this reference), i.e. 12 binding energies
of spherical nuclei distributed all over the periodic table and 9 charge radii.
Due to the fact that the density dependence in the isovector channel
is determined by the equation of state of neutrons it was not necessary
to include neutron skin thicknesses ($r_{n}-r_{p}$) data.
This fit provides us with a relatively stable starting point for a subsequent
fine tuning of the model. Moreover, since the $\delta$-meson is
little influenced by the overall fit to finite nuclei both $g_{\delta}(\rho_{\text{sat}})$ and $f_{\delta}(x)$ (4 parameters) are relatively well determined already in
this step and we need only a fine tuning of the remaining parameters in the
next two steps.

\item[2)] In the second step we keep the four meson masses and the
four parameters describing the density dependent vertex $g_{\delta}(\rho)$
of the $\delta$-meson fixed and determine the 9 parameters describing the
Typel-Wolter ansatz for the density dependent vertices $g_{i}(\rho)$ of
the remaining three mesons ($i=\sigma$, $\omega$ and $\rho$) by a very
accurate fit to the nuclear matter data shown in Figs.~\ref{fig1} and~\ref{fig2}.
Involving only nuclear matter data, this is a relatively fast calculation and
as a result we obtain the three density dependent vertices $g_{i}(\rho)$ for
$i=\sigma$, $\omega$, $\rho$. In this way we describe with high precision the
EoS for symmetric nuclear matter and pure neutron matter as well as the
isovector part of the  effective Dirac mass $\Delta m^\ast_{}=m_{p}^{\ast}-m_{n}^{\ast}$.

\item[3)] In the last step we keep the $\delta$-meson parameters as determined
in step 1 and the density-dependent functions $f_{i}(x)$ for
$i=\sigma, \omega, \rho$ are frozen at the values found in step 2.
We refine the remaining 4 parameters $g_{\sigma}%
(\rho_{\text{sat}})$, $g_{\omega}(\rho_{\text{sat}})$, $g_{\rho}%
(\rho_{\text{sat}}),$ and $m_{\sigma}$ to the binding energies of 161
spherical nuclei and the charge radii of 86 nuclei shown in Table~\ref{tab5}
taking into account in this fit also the pseudo-data of the nuclear matter properties used in step 1 and 2
with certain weights. It turns out that the values of $g_{\sigma}%
(\rho_{\text{sat}})$, $g_{\omega}(\rho_{\text{sat}})$, $g_{\rho}%
(\rho_{\text{sat}}),$ and $m_{\sigma}$ obtained in this fit differ only
slightly from the values determined in step 1 and that the nuclear matter
results (EoS in symmetric nuclear matter and pure neutron matter as well as
the isovector Dirac mass) differ only marginally from the results obtained in
step 2. As a consequence the procedure involving step 2 and 3 does not
have to be repeated.
\end{enumerate}

\begin{table*}[h!]
\begin{center}
\begin{tabular}{lrrrrrrr}
\hline\hline
& & & & & & & \\
$i\quad$   &$m_i$ (MeV)&$g_i(\rho_{\rm sat})$~& $a_i$~~~~& $b_i$~~~~& $c_i$~~~~& $d_i$~~~~& $e_i$~~~~\\
& & & & & & & \\
\hline
& & & & & & & \\
$\sigma$&566.1577 & 10.33254 & 1.392730& 0.1901198 & 0.3678654 & 0.9519078& 0.9519078 \\
$\omega$&783.0000 & 12.29041 & 1.408892& 0.1697977 & 0.3429006 & 0.9859508& 0.9859508 \\
$\delta$&983.0000 & 7.151971 & 1.517787& 0.3262490 & 0.6040782 & 0.4257178& 0.5885143 \\
$\rho  $&763.0000 & 6.312758 & 1.887685& 0.06514596& 0.3468963 & 0.9416816& 0.9736893 \\
& & & & & & & \\
\hline\hline
\end{tabular}
\caption{The parameter set DD-ME$\delta$ with the $\delta$-meson. It
includes 14 independent parameters. Only 4 of them
($g_\sigma(\rho_{\text{sat}})$, $g_\omega(\rho_{\text{sat}})$,
$g_\rho(\rho_{\text{sat}})$, and $m_\sigma$) are fitted to finite
nuclei. The other 10 are derived from an adjustment
to ab-initio calculations in infinite nuclear matter~\cite{BMS.04,DFF.07}}%
\label{tab1}
\end{center}
\end{table*}

The final parameter set DD-ME$\delta$ obtained in this way is given in
Table~\ref{tab1}. It is compared with the parameter set DD-ME2 in
Table~\ref{tab2}. We observe a large difference in the value of the
the $\rho$-nucleon vertex $g_\rho$. This can be understood by the
fact that we have in DD-ME2 only one meson ($\rho$) in the isovector
channel with a very different density dependence. We also observe a
very different density dependence for the isoscalar mesons $\sigma$
and $\omega$. All in all this has only a minor effect in the low
density region $\rho \le \rho_{\text{sat}}$ but it has a large effect
at high densities as shown in Fig.~\ref{fig3}.
\begin{table}[h!]
\begin{center}
\begin{tabular}{lrrrrrrr}
\hline\hline
& & & & & & & \\
$i\quad$   &$m_i$ (MeV)&$g_i(\rho_{\rm sat})$~& $a_i$~~~~& $b_i$~~~~& $c_i$~~~~& $d_i$~~~~& $e_i$~~~~\\
& & & & & & & \\
\hline
& & & & & & & \\
$\sigma$& 550.1238 &10.5396 &1.3881 &1.0943 &1.7057 &0.4421 &0.4421 \\

$\omega$& 783.0000 &13.0189 &1.3892 &0.9240&1.4620 & 0.4775&0.4775 \\
$\rho  $& 763.0000 &~3.6836 &0.5647 &  & & & \\
& & & & & & & \\
\hline\hline
\end{tabular}
\caption{The parameter set DD-ME2 which does not contain a
$\delta$-meson. It includes 8 independent parameters. All are fitted
to finite nuclei. In DD-ME2 the density dependence of the $\rho$-meson
is given by $g_\rho(\rho)=g_\rho(\rho_{\rm sat})\exp(-a_\rho(x-1))$}%
\label{tab2}
\end{center}
\end{table}

\subsection{$\chi^{2}$ definition}
\label{sec3}

Our fit is performed through a $\chi^{2}$ test of the form
\begin{equation}
\chi^{2}=\frac{1}{n_{\mathrm{data}}}\sum_{i=1}^{n_{\mathrm{data}}}w^{2}%
_{i}\left(  {\mathcal{O}}^{\mathrm{model}}_{i}-{\mathcal{O}}^{\mathrm{ref}%
}_{i}\right)  ^{2} \label{chi2}%
\end{equation}
where $n_{\mathrm{data}}$ is the number of data points and $w_{i}$
the weight associated to each data point. ${\mathcal{O}}^{\mathrm{ref}}_{i}$
is the experimental value for finite nuclei and the pseudo data obtained by
ab-initio calculations in nuclear matter. The observables in finite nuclei
used for the fit are the binding energies of $161$ nuclei
and the charge radii of $86$ nuclei given in
Table~\ref{tab5}. All of the isotopes are spherical even-even
nuclei and the data are taken from the literature~\cite{AWT.03}.
In the standard definition of a $\chi^{2}$ test, the weights $w_{i}$
should be inversely proportional to the experimental uncertainties.
However, in the case of energies these are usually so small,
that they cannot be used as relevant quantities. We therefore
used the weights given in Table~\ref{tab3}, i.e. $w_i=1/0.5$ MeV$^{-1}$
for the masses and $w_i=1/0.01$ fm$^{-1}$ for the radii.
For the fit to the results of ab-initio calculations in nuclear
matter we use $n_{\rm data}$ mesh points in a certain density range
(see Table~\ref{tab3}) and we assume a relative accuracy of 3~\%.
The minimization of $\chi^2$ is carried out by means of
a variable metric method algorithm included in the \textsc{MINUIT}
package of Ref.~\cite{Mi96}.

\begin{table}[h!]
\begin{center}
\begin{tabular}{lccccc}
 & & & & & \\
\hline\hline
 & & & & & \\
${\mathcal O_i}$        & $w_i$                 &$n_{\rm data}$&$\rho$ (fm$^{-3}$)&$\chi^2$&Ref. \\
 & & & & & \\
\hline
 & & & & & \\
$B$                  &$1/0.50$ MeV$^{-1}$           &$161$        &                   &23.40 &~\cite{AWT.03}\\
$r_{\rm c}$          &$1/0.01$  fm$^{-1}$           &$86$         &                   &2.90  &~\cite{An04}\\
 & & & & & \\
\hline
 & & & & & \\
$e(\rho,\alpha=1)$  &$1/(0.03\times{\mathcal O}_i)$&$30$         &$0.01 - 0.30$      &3.42  &~\cite{BMS.04}\\
$e(\rho,\alpha =0)$  &$1/(0.03\times{\mathcal O}_i)$&$30$         &$0.01 - 0.30$      &7.03  &~\cite{BMS.04}\\
$m^*_p-m^*_n$        &$1/(0.03\times{\mathcal O}_i)$&$25$         &$0.04 - 0.20$      &0.39  &~\cite{DFF.07}\\
 & & & & & \\
\hline\hline
\end{tabular}
\caption{Specifications of the $\chi^2$ definition. The total $\chi^{2}$ found for DD-ME$\delta$ is almost $40$, the
partial contributions to it are listed in the fifth column.}
\label{tab3}
\end{center}
\end{table}

In the first step of the fit discussed in the last section
we minimize the quantity
\begin{equation}
\chi^2_{(1)}=\chi^2_B+\chi^{2}_{r_c}+\chi ^{2}_{\mathrm{sym}}
+\chi^2_\mathrm{neut}+\chi^2_{\Delta\mathrm{m^*}}%
\label{chi2-1}
\end{equation}
At this stage all the 14 parameters of the model
are varied and the data of finite nuclei
are restricted to the masses and charge radii of the 12 nuclei used
in the fit of the parameter set DD-ME2 in Ref.~\cite{DD-ME2}.
As we have described in the last section, the parameters for the
$\delta$-meson obtained from this fit are no longer changed. We given
them in the third line of Table~\ref{tab1}.

In the second step we minimize
\begin{equation}
\chi^2_{(2)}=\chi ^{2}_{\mathrm{sym}}
+\chi^2_\mathrm{neut}+\chi^2_{\Delta\mathrm{m^*}}%
\label{chi2-2}
\end{equation}
for nuclear matter data with respect to the 6 constants
characterizing the density dependence of the $\sigma$, $\omega$ and
$\rho$ mesons and the 3 couplings $g_{\sigma}(\rho_{\rm sat})$,
$g_{\omega}(\rho_{\rm sat})$, $g_{\rho}(\rho_{\rm sat})$. (Here, the
$\delta$-meson and $m_\sigma$ are hold at the
values found in the first step.) The obtained values for the $a_i, b_i,
c_i, d_i, e_i$ constants that define the density dependence of the
$i=\sigma,\omega,\rho$ meson-nucleon vertices are given in
Table~\ref{tab1}.

In the third step we minimize $\chi^{2}$ for the nuclear matter data
and the $161$ binding energies and $86$ charge radii given in
Table~\ref{tab5}:
\begin{equation} \chi^{2}_{(3)}=\chi^{2}_{B}+\chi^{2}_{r_c}
+\chi ^{2}_{\mathrm{sym}}
+\chi^{2}_{\mathrm{neut}}+\chi^{2}_{_{\Delta
\mathrm{m^{*}}}}%
\label{chi2-3}%
\end{equation}
Now we fit only a restricted set of 4 parameters, i.e., the 3 couplings
$g_{\sigma}(\rho_{\rm sat})$, $g_\omega(\rho_{\rm sat})$,
$g_{\rho}(\rho_{\rm sat})$, and $m_\sigma$. The resulting values
are given in the first two columns of Table~\ref{tab1}.

We have to emphasize that only \textit{four} free parameters
$g_{\sigma}(\rho_{\rm sat})$, $g_{\omega}(\rho_{\rm sat})$, $g_{\rho}(\rho_{\rm sat})$
and $m_{\sigma}$ have been used in the final fit to the experimental
data in finite nuclei. The other 10 parameters are derived from ab-initio
calculations. This is in contrast to the typical relativistic and
non-relativistic fits of mean-field interactions,
where commonly around $10$ free parameters are adjusted to data in finite
nuclei. It is also worth to remember that adding the $\delta$-meson has improved our
theoretical picture of the nucleus and of the EoS of
asymmetric nuclear matter.

So far we have used in the fit only nuclei with spherical shapes.
The pairing correlations are treated in the first step in the constant gap approximation
with gap parameters derived from the odd-even mass differences. For a full description
of nuclei all over the periodic table which includes also regions where the experimental binding energies are not known,
we introduce a more general description of pairing by means of a monopole force
with a constant matrix element fitted to reproduce the experimental binding energies of the nuclei in Table~\ref{tab5}.
We obtain for the set DD-ME$\delta$ the values $G_n = 32.44/A$ MeV for neutrons and $G_p = 29.76/A$ MeV for protons.
In order to have a fair comparison for the results in finite nuclei we treated in the following
the pairing properties of the set DD-ME2 also by a monopole force. In a similar fit we found for
DD-ME2 the strength parameters $G_n = 29.86/A$ MeV and $G_p = 28.92/A$ MeV.
In all these calculations the soft pairing window described in Ref.~\cite{BRR.00} has been used.


\section{Results}

\label{results}

\subsection{Nuclear and Neutron Matter Equations of State}

\begin{table}[b]
\begin{center}
\begin{tabular}{lr@{.}lr@{.}lr}
\hline\hline\\
 &\multicolumn{2}{c}{DD-ME$\delta$} & \multicolumn{2}{c}{DD-ME2} & \\%
\\
\hline\\
$\rho_{\rm sat}$& ~~~~~~~0&152   &~~~~~~0&152 &~~~[fm$^{-3}$] \\
$e_{\rm sat}$   &    $-$16&12    &  $-$16&14  &   [MeV]       \\
$K_\infty$      &      219&1     &    250&89  &   [MeV]      \\
$J$             &     ~~32&35    &   ~~32&30  &   [MeV]      \\
$L$             &     ~~52&85    &   ~~51&26  &   [MeV]      \\
$m^*/m$         &        0&609   &      0&572 &              \\
\\
\hline\hline
\end{tabular}
\caption{Nuclear saturation properties as predicted by the parameter
sets DD-ME$\delta$ and DD-ME2}%
\label{tab4}
\end{center}
\end{table}

The nuclear matter properties at saturation computed with the DD-ME$\delta$ functional
are given in Table~\ref{tab4}. These properties do not fully coincide with
the ones of the fully microscopic calculation in ~\cite{BMS.04}.
The reason for that is that in the microscopic calculation, the EoS is very
flat around saturation density and some deviation between the
microscopic results and the DD-ME$\delta$ fit appear. These differences remain
within the uncertainty of the state of the art of present numerical microscopic calculations.
They are too small to be seen on the scale of Fig.~\ref{fig1}.
They are, however, important for a fine tuning of the results.



\begin{figure}[t]
\vspace{0.6cm}
\begin{center}
\includegraphics[width=8cm,clip=true]{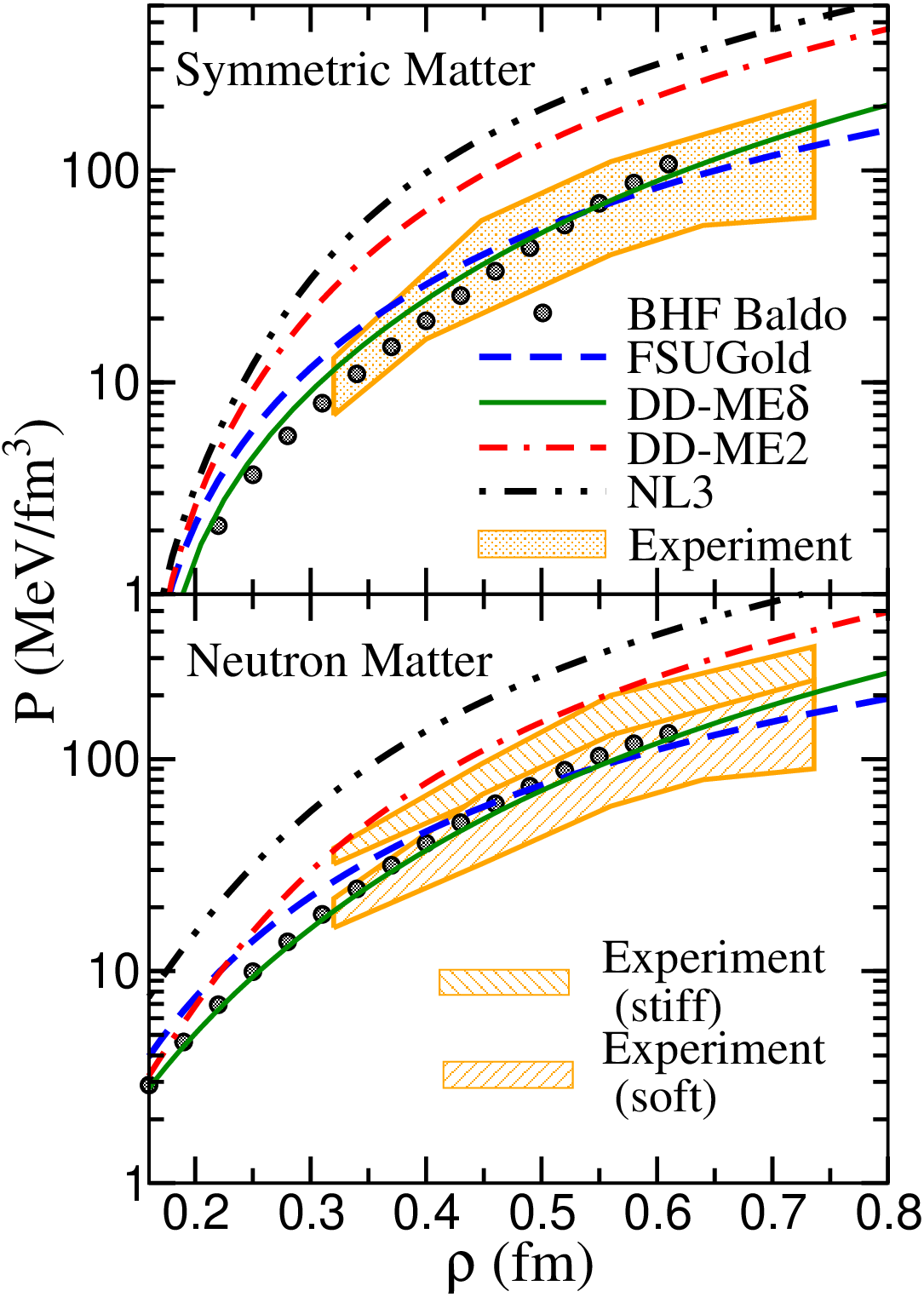}
\end{center}
\vspace{-0.5cm}%
\caption{(Color online) The pressure defined in Eq.(\ref{pressure}) for symmetric nuclear matter (upper panel) and
for neutron matter (lower panel) as a function of the density.
The results for DD-ME$\delta$ are compared with those of the density functionals
NL3 from Ref.~\cite{NL3}, %
FSUGold from Ref.~\cite{TP.05}, %
DD-ME2 from Ref.~\cite{DD-ME2}, %
and microscopic
BHF calculations from Ref.~\cite{BMS.04}. %
The shaded area represents experimental results
from Ref.~\cite{DLL.02}.}%
\label{fig3}
\end{figure}

In order to investigate the quality of the predictions of the density functional DD-ME$\delta$ in the high density
domain, we show in Fig.~\ref{fig3} the pressure (\ref{pressure}) computed with this functional as a function
of the density. It is compared with the pressure derived from the microscopic calculation of Ref.~\cite{BMS.04}
as well as with the results derived from the non-linear meson coupling models NL3~\cite{NL3} and FSUGold~\cite{TP.05}
and from DD-ME2~\cite{DD-ME2}. We see that both, microscopic and DD-ME$\delta$ calculations, are within the shaded
area which corresponds to the "experimental region" estimated from simulations of heavy-ion collisions~\cite{DLL.02}.
The standard non-linear $\sigma$-$\omega$ model NL3 is outside of this region while the FSUGold model -- which has
an additional non-linear $\omega$-$\rho$ coupling that softens the symmetry energy (see below) -- is
inside the shaded area in rather good agreement with DD-ME$\delta$ and the microscopic results. The results of the
parameter set DD-ME2 are slightly outside of the shaded area.

An important quantity in nuclear physics and astrophysics, directly related with the EoS of asymmetric nuclear matter,
is the symmetry energy (\ref{csym}). The value of the symmetry energy derived from successful mean field
models lies roughly in a window of $30-35$ MeV at saturation. However, the density dependence of the symmetry
energy is much more uncertain. This fact entails important consequences for a number of isospin-dependent observables.
As a paradigmatic example, one may recall that different accurate mean-field models which reproduce well the binding
energy and charge radius of the nucleus $^{208}$Pb predict largely different values for the neutron skin thickness of
this isotope, ranging from $0.1-0.3$ fm. This fact points out that the isovector properties of the different models are,
actually, not well constrained by the binding energies and charge radii of stable finite nuclei used to fit the
effective interactions.

In nuclear mean field models, a strong linear correlation exists~\cite{Bro.00,RCV.11} between the size
of the neutron skin thickness of a heavy neutron-rich nucleus such as $^{208}$Pb
and the $L$ parameter defined in Eq. (\ref{esymmexp}), i.e., the slope of the symmetry energy at saturation. Recent
constraints on the L parameter have been obtained using a variety of observables such as, for instance, isospin
diffusion~\cite{SSK.06,SYS.07a,SYS.07b} and isoscaling ~\cite{SL.05,CKL.05a,CKL.05b,CKL.07,LCK.08} in heavy ion
reactions, some collective excitations in nuclei~\cite{VNR.03,AKF.05,KPA.07,TCV.08} and the neutron skin thickness in
finite nuclei~\cite{WVR.09,CRV.09} measured in antiprotonic atoms~\cite{TJL.01,KRJ.07}. The analysis
of all these results suggests that the $L$ parameter is roughly within the window 45 - 75 MeV~\cite{WVR.09}. The new
experimental efforts to measure the neutron radius of $^{208}$Pb may turn out to be helpful to deduce in the future
narrower constraints on the slope L of the symmetry energy through the correlation of $L$ with the neutron skin
thickness~\cite{ZSM.10,RCV.11}.
\begin{figure}[h]
\vspace{0.6cm}
\begin{center}
\includegraphics[width=8cm,clip=true]{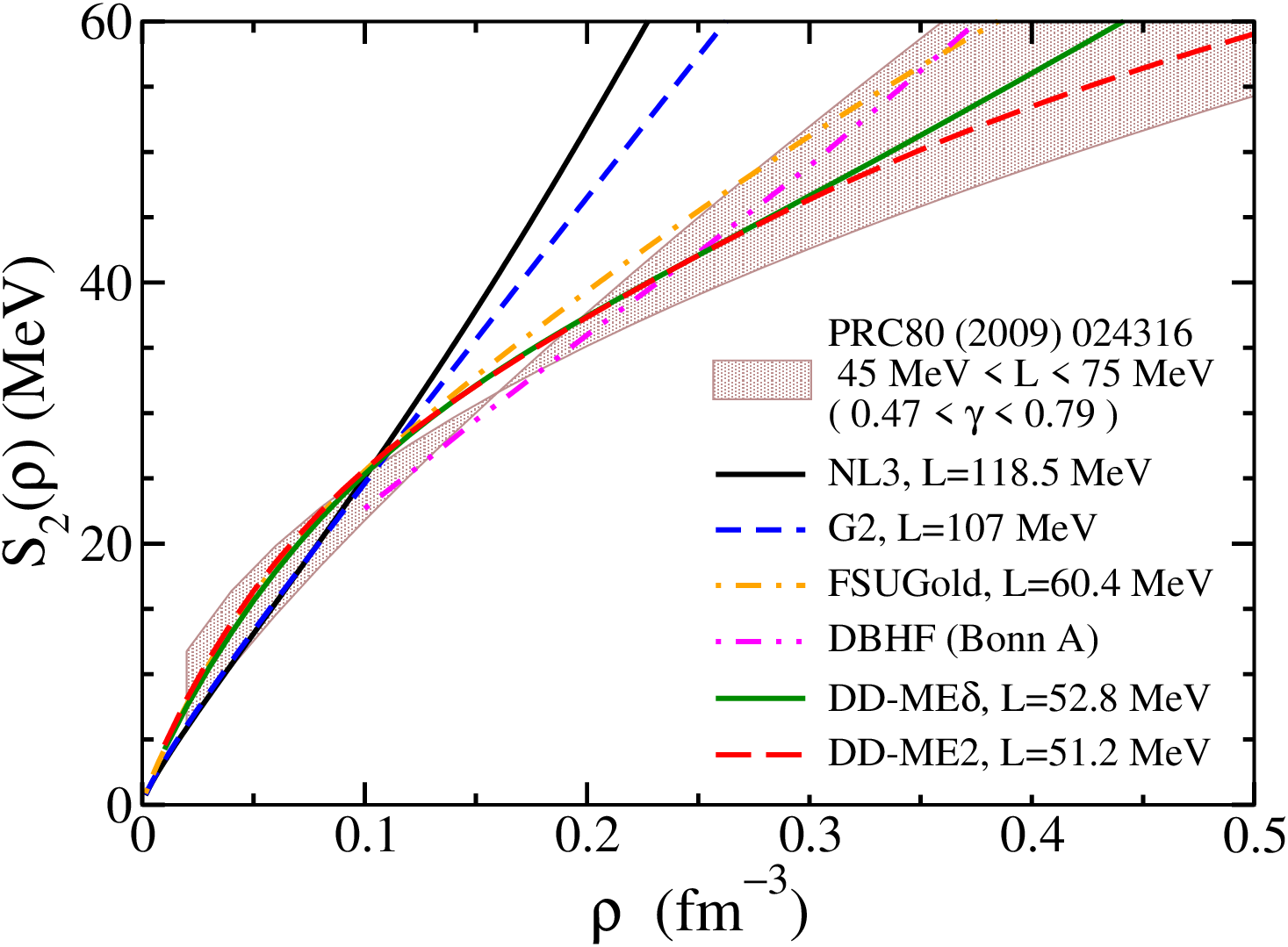}
\end{center}
\vspace{-0.5cm}%
\caption{(Color online) The symmetry energy in Eq. (\ref{csymdef2}) as a function of the density.
The results for DD-ME$\delta$ are compared with those of the density functionals
NL3 from Ref.~\cite{NL3}, %
G2 from Ref.~\cite{Fu97}, %
FSUGold from Ref.~\cite{TP.05}, and microscopic
DBHF calculations from Ref.~\cite{LMB.92}. %
The shaded area represents the empirical region suggested
by the available constraints on the $L$ parameter
discussed in~\cite{WVR.09}.}%
\label{fig4}%
\end{figure} 

The $L$ value predicted by our novel DD-ME$\delta$ functional is $53$ MeV,
close to the result of the microscopic calculation of $66.5$ MeV in Ref.~\cite{VPP.09}.
The density dependence of the symmetry energy exhibited
by DD-ME$\delta$, is displayed in Fig.~\ref{fig4}. We
see that DD-ME$\delta$ predicts a rather soft
density dependence of the symmetry energy which lies inside the
shaded region derived from the empirical law
$S_2(\rho)=31.6(\rho/\rho_{0})^{\gamma}$ MeV imposing the range
discussed above: $45$ MeV $<L<$ $75$ MeV. It turns out that the density dependence of DD-ME$\delta$ and DD-ME2
is practically the same. This fact is not trivial, first because DD-ME2
has not been adjusted to nuclear matter data, but only to the experimental
skin thickness of several finite nuclei~\cite{DD-ME2} and second, the full isospin
dependence is determined by the $\rho$-meson, whereas in DD-ME$\delta$ it is distributed over the
$\delta$ and the $\rho$-meson. 

\begin{figure}[h]
\vspace{0.6cm}
\begin{center}
\includegraphics[width=8cm,clip=true]{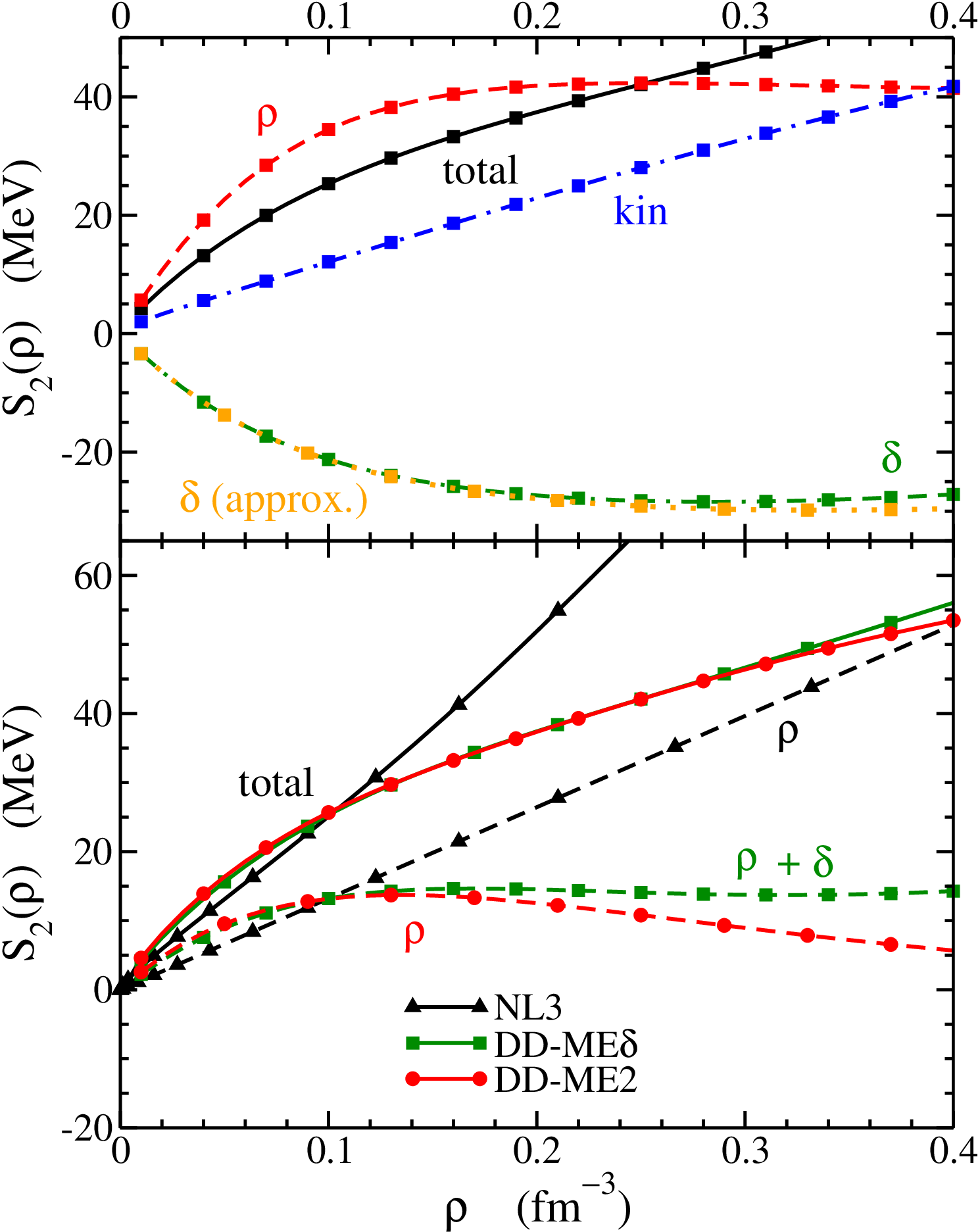}
\end{center}
\caption{(Color online) Upper panel: The symmetry energy $S_2(\rho)$ (full black) and its contributions
as a function of the density, the kinetic contribution (dash dotted in blue), the contributions
of the $\rho$-meson (dashed in red), the $\delta$-meson (dash dotted in green), and
approximation~(\ref{cdapp}) for the contribution of the $\delta$-meson (dotted in yellow).
Lower panel: The symmetry energy $S_2(\rho)$
resulting from the parameter sets NL3 (triangles), DD-ME2 (circles),
and DD-ME$\delta$ (squares). The total value (solid line) is compared with
the contribution of the $\rho$ meson (dashed line) for NL3 and DD-ME2
and of the sum of $\rho$ and $\delta$ mesons for DD-ME$\delta$.}%
\label{fig5}%
\end{figure}

The reason for this good agreement can be understood from the upper
panel of Fig.~\ref{fig5}  where the different contributions to the symmetry energy
are displayed, the kinetic part as well as those provided by the $\rho$ and the
$\delta$-meson. We can see that the contributions of these mesons have opposite
sign and thus a noticeable cancellation appears between them over the
entire range of densities under consideration. Thus, it is conceivable (see
Eq. (\ref{E39})) that if the $\delta$-meson is not considered in the functional 
(as it is the case of DD-ME2) its contribution to the symmetry energy can be 
accounted for by the $\rho$-meson (with a reduced strength of the coupling 
constant, see Ref. \cite{DD-ME2} and Table~\ref{tab2}).

The lower panel shows similar decompositions of the symmetry energy for
other density functionals, such as NL3 \cite{NL3} and DD-ME2 \cite{DD-ME2}. 
The parameter set NL3 (black) has no density dependence in the isovector channel.
Therefore the contribution of the $\rho$-meson is very stiff and proportional to
the density. The parameter set DD-ME2 includes only one isovector meson, 
the $\rho$-meson and its contribution to the symmetry energy is very close
the the sum of both the $\rho$- and the $\delta$-meson for the set DD-ME$\delta$ which
compensate each other to a large extend. Small differences in these curves
at densities above saturation density can be traced back to the different
ansatz for the density dependence of the $\rho$-meson in these two parameter
sets, the Typel-Wolter ansatz (\ref{fx}) for DD-ME$\delta$ and an exponential density
dependence for DD-ME2 (see Eq. (7) in Ref.~\cite{DD-ME2}).

\subsection{Ground-state properties of finite nuclei}

\begin{figure}[b]
\vspace{0.6cm}
\begin{center}
\includegraphics[width=8cm,clip=true]{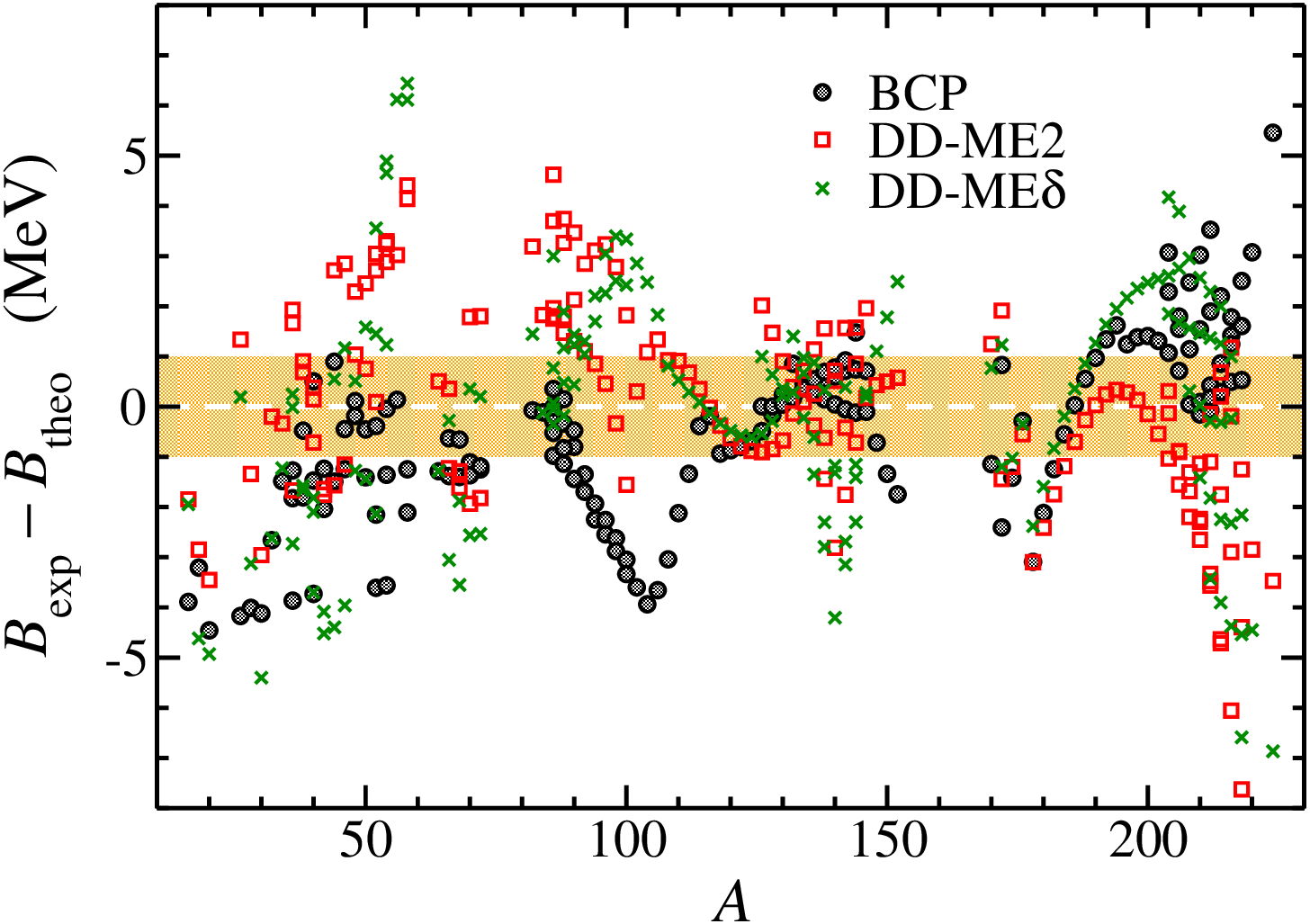}
\end{center}
\vspace{-0.5cm}%
\caption{(Color online) Difference between theoretical and experimental binding
energies as a function of the mass number. Results of the functional DD-ME$\delta$
are compared with those of DD-ME2~\cite{DD-ME2} and of BCP~\cite{BSV.08}.
The orange region corresponds to twice the fixed weight used in the fit (see Table
\ref{tab3}).}%
\label{fig6}%
\end{figure}

As described in section~\ref{section3a} the experimental masses of $161$ and the charge
\textit{rms} radii of $86$ even-even spherical nuclei (see Table~\ref{tab5}) have been
taken into account in the fitting procedure of the DD-ME$\delta$ functional.

\begin{figure}[t]
\vspace{0.6cm}
\begin{center}
\includegraphics[width=8cm,clip=true]{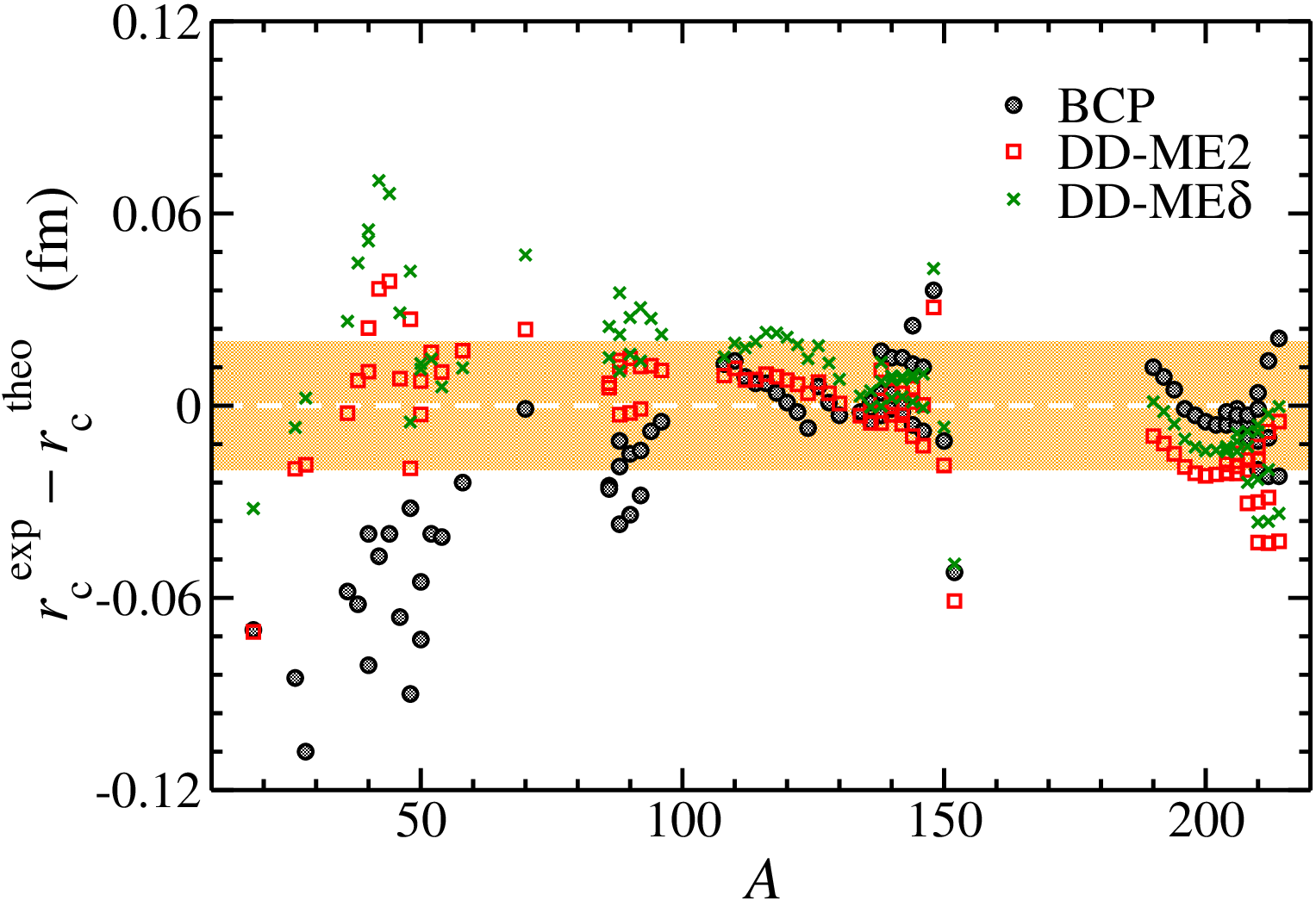}
\end{center}
\vspace{-0.5cm}%
\caption{Color online) Difference between theoretical and experimental charge radii
as a function of the mass number. DD-ME$\delta$ results as compared
with those of DD-ME2~\cite{DD-ME2} and of BCP~\cite{BSV.08}. The orange region
corresponds to twice the fixed weight used in the fit (see Table~\ref{tab3}).}%
\label{fig7}%
\end{figure}

We display in Figs.~\ref{fig6} and~\ref{fig7} the difference between theoretical results computed with the functionals
DD-ME$\delta$, DD-ME2 and BCP and experimental data. For DD-ME$\delta$ we obtain a \textit{rms} deviation of $2.4$ MeV for the binding energies and of $0.02$ fm for the charge radii. These results are close to the \textit{rms} deviations $2.1$ MeV and $0.02$ fm obtained with the DD-ME2 functional for the same set of data when pairing correlations are introduced by the monopole force discussed at
the end of Sect.~\ref{sec3}.
It has to be emphasized, however, that using the density functional DD-ME2 in connection with the pairing part of the finite
range Gogny force D1S instead of the monopole force and taking into account spherical as well as deformed nuclei one has found \textit{rms} deviations of 900 keV and 0.017 fm for the binding energies and charge radii of typical sets of 200~\cite{DD-ME2} or 300~\cite{Lalazissis-private} even-even nuclei.

\begin{figure}[b]
\begin{center}
\includegraphics[width=8cm,clip=true]{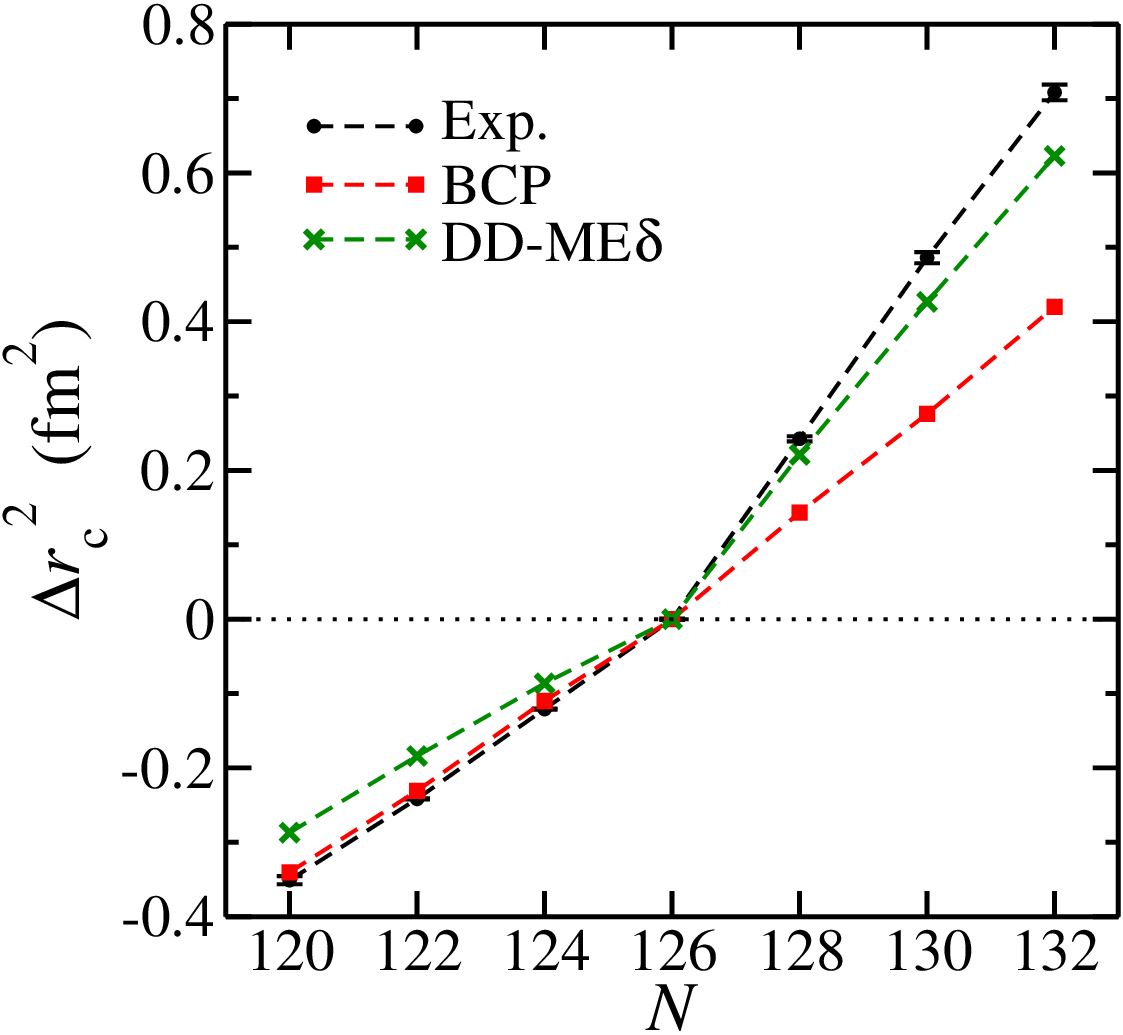}
\end{center}
\vspace{-0.5cm}%
\caption{(Color online) Isotope shifts for the chain of Pb isotopes with respect
to $^{208}$Pb. Calculations with the relativistic model DD-ME$\delta$ and
the non-relativistic functional BCP are compared.}%
\label{fig8}%
\end{figure}

The charge radii $r_c$ (defined in Eq. (\ref{rch})) of Pb isotopes
and their isotope shifts have been a matter of detailed discussion within
the framework of mean field theories~\cite{TBF.93,SLR.93,SLK.95,RF.95}.
In Fig.~\ref{fig8} we show the isotope shifts in a chain of Pb
isotopes as a function of the neutron number $N$. The nucleus $^{208}$Pb
has been taken as the reference point: $\Delta^2_{r_c}(N)=r^2_c(N)-r^2_c(126)$.
With a gradual addition of neutrons, the empirical charge radii of isotopes
heavier than $^{208}$Pb do not show the trend of the lighter isotopes and at
the doubly magic nucleus $^{208}$Pb one observes a pronounced kink~\cite{Ott.89}.
Conventional non-relativistic Skyrme- and Gogny forces
fail to reproduce this kink~\cite{TBF.93}, whereas all the
relativistic models are successful in describing this kink
properly~\cite{SLR.93}. In Refs.~\cite{SLK.95,RF.95} this difference
between the non-relativistic Skyrme functional and the
relativistic models has been traced back to the isospin
dependence of the spin-orbit force. In conventional relativistic
models it is determined by the $\rho$-meson vertex and it is relatively
weak. In Fig.~\ref{fig8} we see that the parameter set DD-ME$\delta$
reproduces the kink in the isotope shifts rather well as all the other
relativistic models do. The non-relativistic set BCP of Ref.~\cite{BSV.08}
which has the same spin-orbit force as conventional Skyrme and Gogny functionals
fails in this context.

\begin{figure}[h]
\vspace{0.6cm}
\begin{center}
\includegraphics[width=8cm,clip=true]{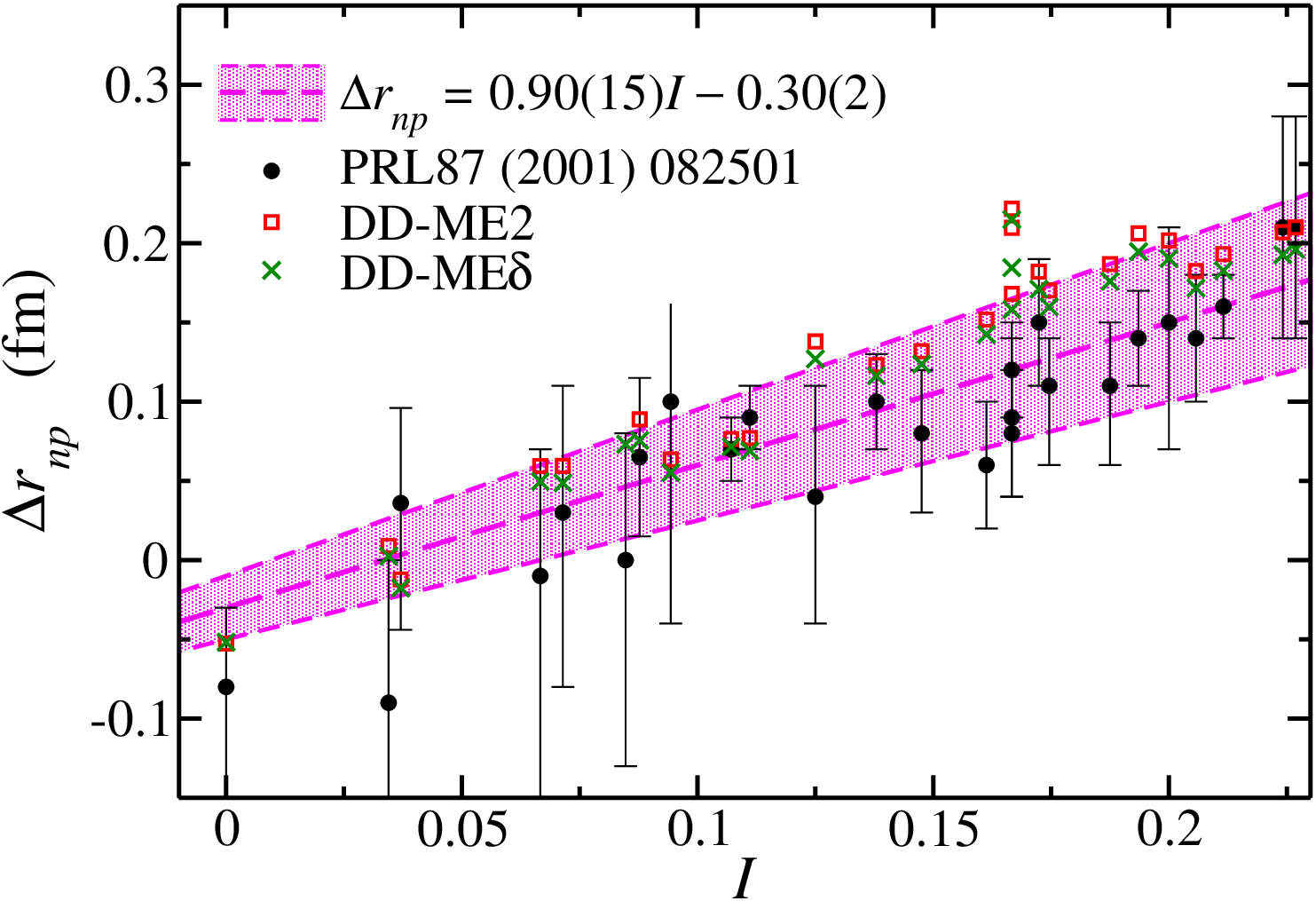}
\end{center}
\vspace{-0.5cm}%
\caption{(Color online) The neutron skin thickness $\Delta r_{np} = \langle r^2\rangle^{1/2}_n-\langle r^2\rangle^{1/2}_p$
as a function of the asymmetry parameter $I=(N-Z)/(N+Z)$. Results obtained with the
parameter set DD-ME$\delta$ are compared with those of the set DD-ME2~\cite{DD-ME2} and
experimental values~\cite{TJL.01}.}%
\label{fig9}%
\end{figure}

Finally we show in Fig.~\ref{fig9} values for the neutron skin thickness
$\Delta r_{np} = \langle r^2\rangle^{1/2}_n-\langle r^2\rangle^{1/2}_p$
of a large set of nuclei as a function of the relative neutron excess $I=(N-Z)/(N+Z)$
and compare the results obtained with the parameter set DD-ME$\delta$ with
those of the set DD-ME2 and with experimental values~\cite{TJL.01}.
Both theoretical calculations are in rather good agreement and within the
range of the experimental error bars.

\subsection{Impact of the $\delta$-meson on the spin-orbit splitting}

In this work, we have included the $\delta$-meson in our theoretical treatment
of the nucleus motivated by microscopic calculations~\cite{HKL.01,SOA.05,DFF.07} and
by the importance of a scalar-isovector meson of the nucleon-nucleon
potentials for describing the nucleon-nucleon scattering data in the
vacuum~\cite{Mac.89}. In our investigation of the properties of nuclear matter
we have seen in Fig.~\ref{fig5} that the influence of the $\delta$-meson on the
symmetry energy can be largely compensated by renormalizing the $\rho$-meson
coupling constant in the DD-ME2 model. The same seems to be true also for the masses
(Fig.~\ref{fig6}), radii (Fig.~\ref{fig7}) and skin thicknesses (Fig.~\ref{fig9})
in finite nuclei. Obviously this also applies for all the other successful covariant
density functionals without the $\delta$-meson degree of freedom.

In order to get a better understanding of these results we follow
Ref.~\cite{KR.91} and eliminate the small components of the spinor $\psi_{i}$
in the Dirac equation (\ref{dirac}). For the large components $f_{i}(\mathbf{r)}$
we are left with a Schroedinger like equation
\begin{equation}
\left\{  \sigma p\frac{1}{2m+\varepsilon_{i}+V_{-}}\sigma p+V_{+}\right\}
f_{i}=\varepsilon_{i}f_{i}.%
\end{equation}
It contains the potentials%
\begin{equation}
V_{\pm}=\Sigma^{s}\pm\Sigma^{0}%
\end{equation}
The potential $V_{+}\approx-50$ MeV corresponds to the conventional potential
in the corresponding non-relativistic Schr\"{o}dinger equation. In theories containing
the $\rho$ and the $\delta$-mesons it can be decomposed into an isoscalar and an isovector part
\begin{equation}
V_{+}(r)= V_+^{IS}(r)+\tau_{3}V_+^{IV}(r)
\end{equation}
with
\begin{align}
V_+^{IS}(r) &= g^{}_\sigma \sigma(r) + g^{}_\omega\omega^0_{}(r)  \\
V_+^{IV}(r) &=  g^{}_\delta \delta^{}_3(r) + g^{}_\rho\rho^0_3(r)  ,
\end{align}
where $\sigma(r)$, $\omega^0_{}(r)$, $\delta^{}_3(r)$, and $\rho^0_3(r)$ are the corresponding meson fields.
In theories without $\delta$-meson the $\rho$-meson vertex has to be renormalized and we find
for the isovector part a pure $\rho$-field
\begin{equation}
V_+^{IV}(r) =  \tilde{g}^{}_\rho\rho^0_3(r)  ,
\end{equation}
with a renormalized coupling $\tilde{g}_\rho$. Since the field $\delta^{}_3(r)$ and $\rho^0_3(r)$ have opposite
sign the renormalized coupling $\tilde{g}_\rho$ has to be considerably smaller than the original $g_\rho$,
as it is seen in Tables~\ref{tab1} and ~\ref{tab2}.

\begin{figure}[h]
\vspace{0.6cm}
\begin{center}
\includegraphics[width=8cm,clip=true]{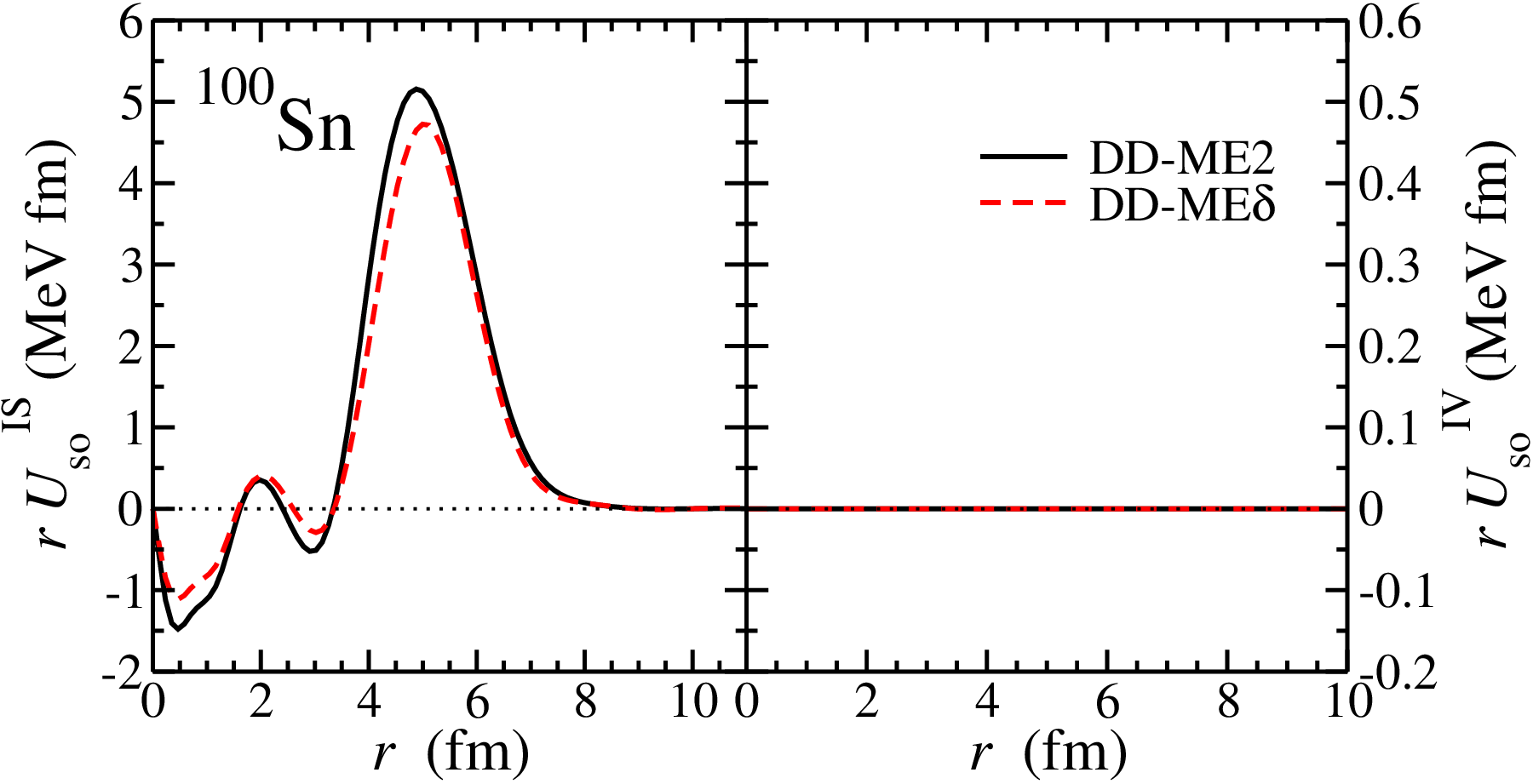}
\vspace{1.0cm}
\includegraphics[width=8cm,clip=true]{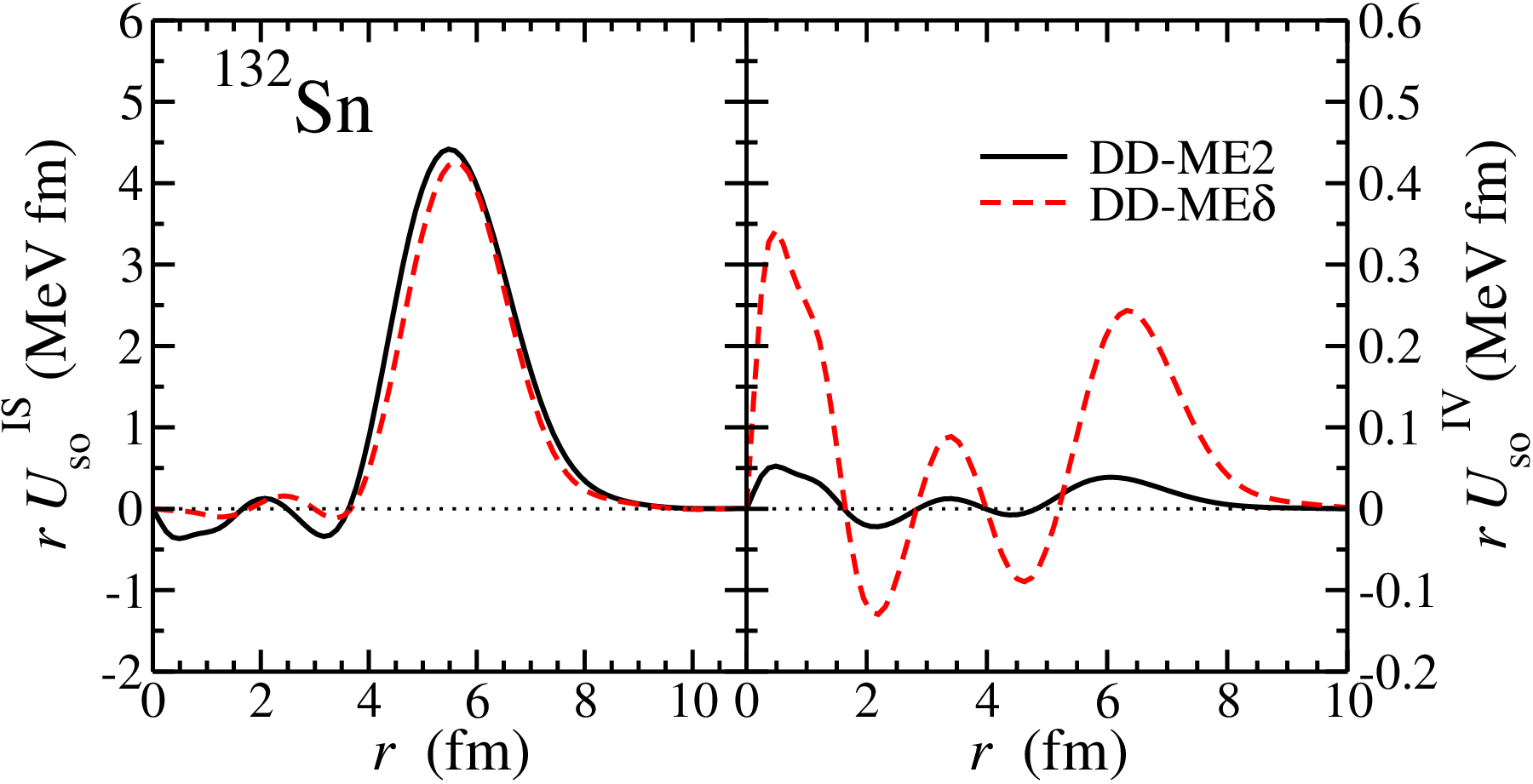}
\vspace{1.0cm}
\includegraphics[width=8cm,clip=true]{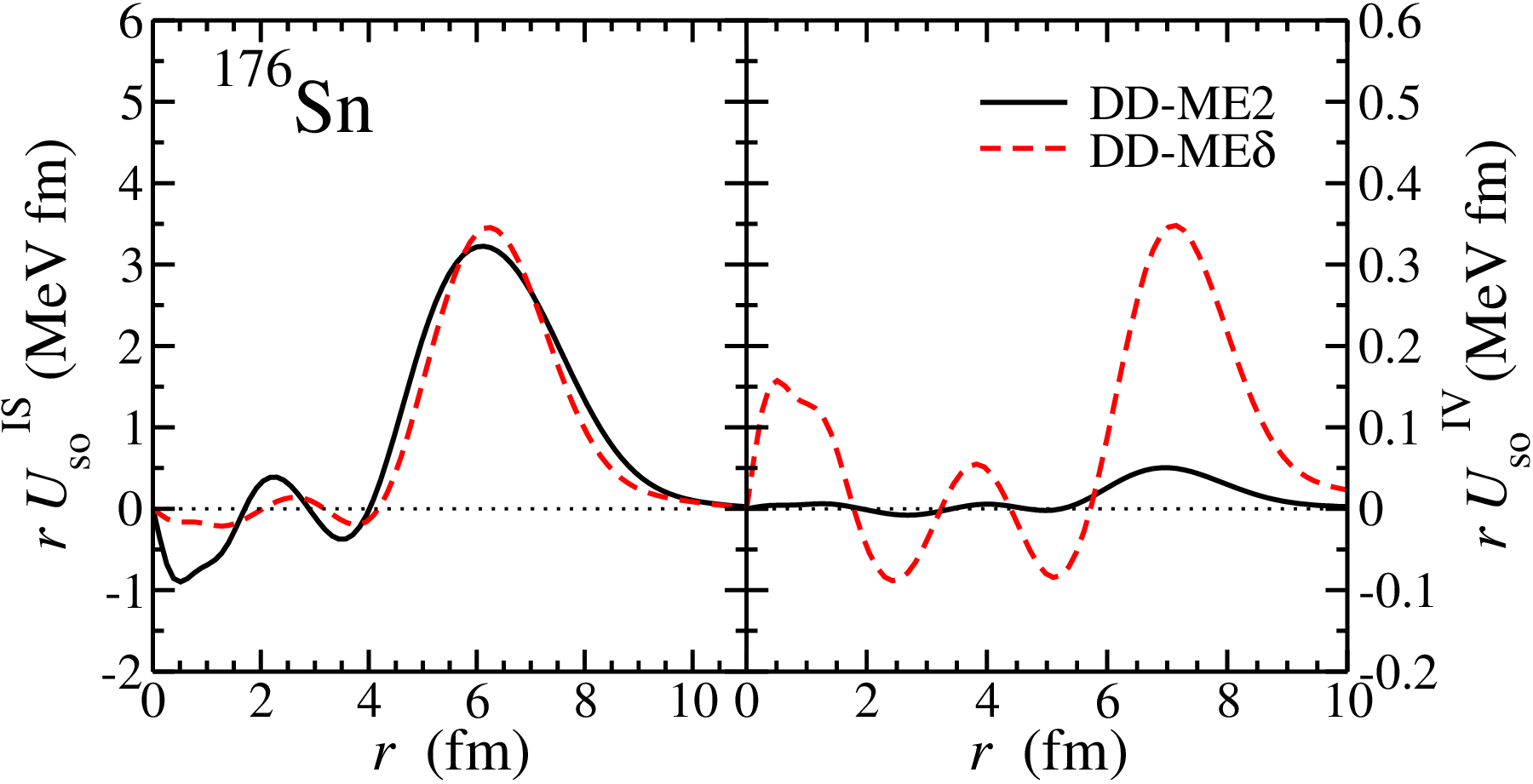}
\end{center}
\vspace{-0.5cm}%
\caption{(Color online) The isoscalar part (left panel) and the isovector part (right panel)
of the spin-orbit potential for the Sn-isotopes $^{100}$Sn, $^{132}$Sn, and $^{176}$Sn.}%
\label{fig10}%
\end{figure}

The situation is different for the potential $V_{-}\approx -700$ MeV, which leads to a very large spin-orbit
term. In spherical nuclei it has the strength $U_{\tau}^{\mathrm{so}}(r)$ for neutrons and protons
($\tau=n,p$) of the form%
\begin{equation}
U^{\tau}_{\mathrm{so}}(r)\equiv\frac{1}{2m}\frac{1}{2m+V_{-}^{\tau}}%
\frac{1}{r}\frac{\partial V_{-}^{\tau}}{\partial r}
\end{equation}
Because of the non-linear connection between $U_{\mathrm{so}}$ and $V_{-}$ the situation is more complicated here
and we obtain the isoscalar and isovector parts as
\begin{align}
U^{IS}_{\mathrm{so}}&=\frac{1}{2}\left(U^{n}_{\mathrm{so}}+U^{p}_{\mathrm{so}}\right) \label{usoisoscal}\\
U^{IV}_{\mathrm{so}}&=\frac{1}{2}\left(U^{n}_{\mathrm{so}}-U^{p}_{\mathrm{so}}\right)  , \label{usoisovec}%
\end{align}
Nonetheless these terms are dominated by $V_-(r)$ and the decomposition of this function with respect to isospin is
\begin{align}
V_-^{IS}(r) &= g^{}_\sigma\sigma(r) - g^{}_\omega\omega^0_{}(r)  \\
V_-^{IV}(r) &= g^{}_\delta\delta^{}_3(r) - g^{}_\rho \rho^0_3(r)  ,
\end{align}
Because of the opposite sign of the fields $\sigma$ and $\omega$ the corresponding isoscalar part of the
spin-orbit potential is considerably enhanced with respect to the isoscalar part of the normal potential.
This well known fact is also true for the isovector part. It is also considerably enhanced with respect to
the isovector part of the normal potential. As a consequence there is an essential difference between a
theory with and  without a $\delta$-meson, i.e. we expect an enhancement of the isospin-dependence
of the spin-orbit potential in a theory with a $\delta$-meson.

To clarify these statements we show in Fig.~\ref{fig10} the isoscalar  and the isovector
part of the spin-orbit potential defined in Eqs. (\ref{usoisoscal}) and (\ref{usoisovec}) for
the three Sn-isotopes $^{100}$Sn, $^{132}$Sn, and $^{176}$Sn. In order to have a clear evidence for the isospin
dependence we neglect in this case the Coulomb potential.

In the first panel ($^{100}$Sn) we have $N=Z$ and because of neglecting the Coulomb potential the neutron and
proton densities are identical. As a consequence the isovector meson fields $\delta(r)$ and $\rho(r)$, as
well as the isovector part of the spin-orbit potential vanish identically. This is true for both models
DD-ME$\delta$ and DD-ME2. Because of the different coupling constants in these models the isoscalar parts
of the spin-orbit potentials are slightly different, but large in both cases and peaked at the surface.
In the second panel ($^{132}$Sn) we have a considerable neutron excess. The isoscalar part did not change
very much. Apart from the fact that the larger mass number $A=132$ produces a shift of the surface and
the maximum of the spin-orbit potential to larger $r$-values, both models show similar results.
The situation is very different for the isovector part. For the DD-ME2 model without $\delta$-meson the
$\rho$-field is not vanishing but relatively small as compared to the isoscalar fields (please consider the
change of the scale in the right hand side of this figure). The effective coupling for the $\rho$-exchange value
of $g^2_\rho$ is considerably smaller than the other two couplings for the other two mesons $\sigma$ or $\omega$.
In addition the source of the $\rho$ field is the difference between the neutron and the proton density.
This difference is even for N=82 not extremely large. On the other side for the DD-ME$\delta$ model with a
$\delta$ and a $\rho$ meson the isovector part of the spin-orbit field is considerably enhanced with respect to
the isovector part of the DD-ME2 model. However, since this is relatively small, in total the isovector part
of the spin-orbit potential is still an order of magnitude smaller than the isoscalar part. Therefore even
for effects, which depend on the spin-orbit potential we do not expect essential differences between
a model with and without a $\delta$-meson. This is even true for cases with extreme neutron excess
as in the nucleus $^{176}$Sn in the lowest panel of Fig.~\ref{fig10}.

\begin{figure}[t]
\begin{center}
\includegraphics[width=8cm,clip=true]{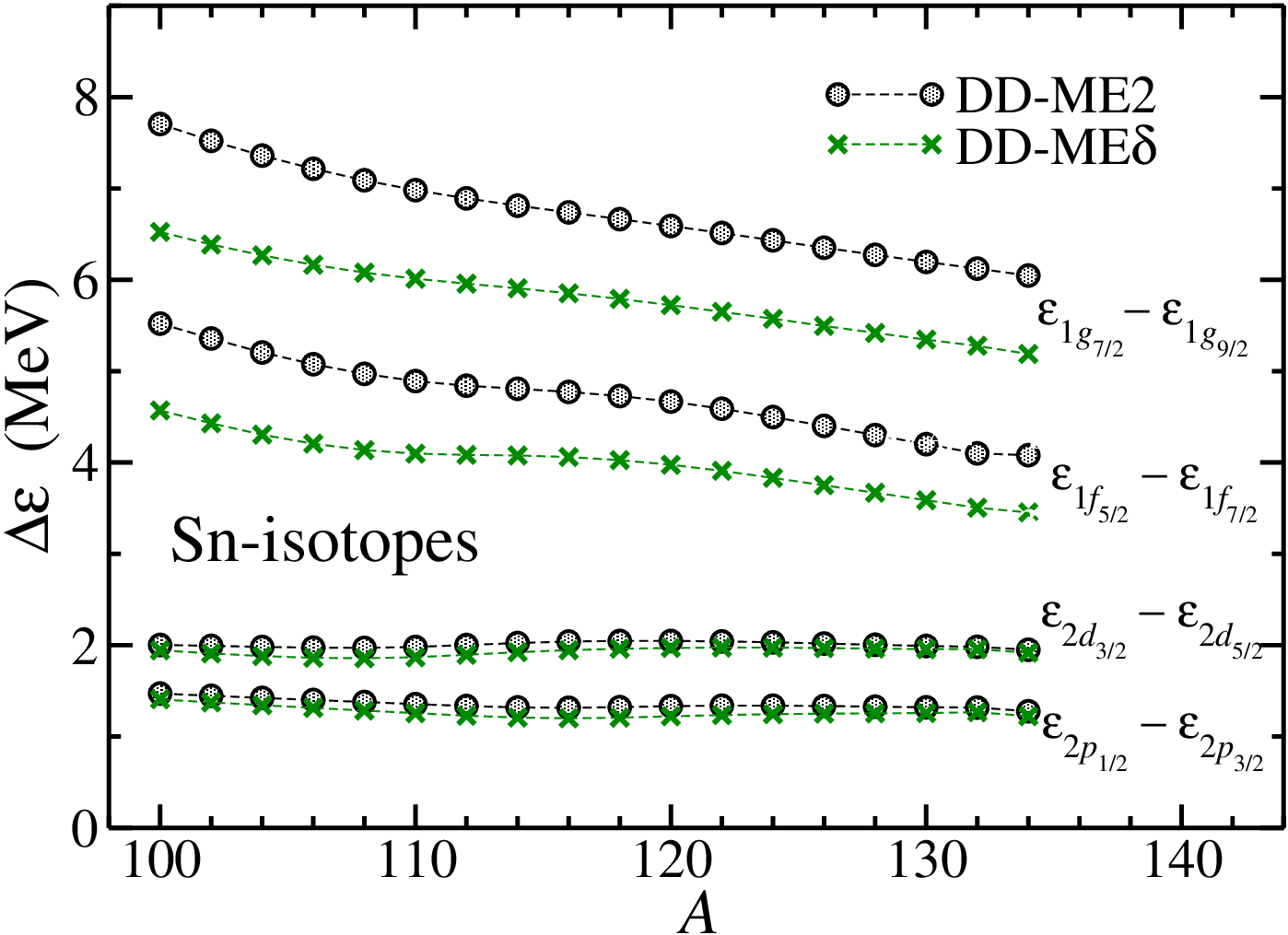}
\end{center}
\vspace{-0.5cm}%
\caption{(Color online) Spin-orbit splitting in Sn-isotopes for various
$nl$ levels. Results for the parameter set DD-ME$\delta$ (green crosses)
are compared with those of the parameter set DD-ME2 (black circles) }%
\label{fig11}%
\end{figure}

In Fig.~\ref{fig11} we show the spin-orbit splitting for neutron orbitals in
a chain of Sn-isotopes, starting at the $N=Z$ nucleus $^{100}$Sn. In contrast to
Fig.~\ref{fig10} the Coulomb interaction is included here. Results for the two
parameter sets DD-ME2 (without $\delta$-meson) and DD-ME$\delta$ (with $\delta$-meson)
are compared. First we find that the difference of these two models is rather small
for the $2p$- and the $2d$-orbits. Because of the low $\ell$-values these
splittings are relatively small and the corresponding wave functions are not so surface peaked.
For the $1f$- and the $1g$-orbits with large $\ell$-values the splitting is relatively
large and we find a considerable difference between the DD-ME2 and the DD-ME$\delta$ model.
This difference is, however, connected with the isoscalar part of the spin-orbit potential,
because it occurs already in the nucleus $^{100}$Sn which has, apart from a small violation
of isospin due to the Coulomb force, practically no isovector part.

With increasing neutron number the spin-orbit splitting in these high $\ell$ orbitals
decreases. This has already been observed in earlier investigations in Ref.~\cite{LVR.98}
and~\cite{LVP.98}, where it has been explained by the increasing neutron skin and the
increasing neutron diffuseness leading to a reduced derivative in the spin-orbit potential.

Finally we observe in Fig.~\ref{fig11} that the difference in the neutron spin-orbit splittings
calculated with DD-ME2 and DD-ME$\delta$ decreases slightly with increasing neutron number.
This effect has its origin in the increasing isovector part of the neutron spin-orbit splitting
for the parameter set DD-ME$\delta$: $U^{n}_{\mathrm{so}}=(U^{IS}_{\mathrm{so}}+U^{IV}_{\mathrm{so}})/2$.
However, it is relatively small because the isovector part itself is small as compared to the
isoscalar part (see the scales on the right panels in Fig.~\ref{fig10}. Of course, for the
protons with $U^{p}_{\mathrm{so}}=(U^{IS}_{\mathrm{so}}-U^{IV}_{\mathrm{so}})/2$ this
difference increases with increasing neutron number (not shown in Fig.~\ref{fig11}).

\section{Discussion and Conclusions}

On the way to a more microscopic derivation of relativistic nuclear energy density
functionals, we started with Brueckner calculations~\cite{BMS.04} for symmetric nuclear matter
and pure neutron matter and with Dirac Brueckner calculations ~\cite{DFF.07} for pure neutron matter.
We tried to use this microscopic information as far as possible for the adjustment of a new covariant
density functional based on density dependent meson exchange. Since it is well known that,
at present, all attempts to derive the functionals directly from bare forces do not reach
the required accuracy for nuclear structure applications we added experimental data in finite nuclei,
such as binding energies and charge radii for the fit. In contrast to Ref.~\cite{DD-PC1}, where a similar
idea has been applied to a relativistic point coupling model, we took into account in this work the
fact that in Dirac Brueckner calculations the resulting scalar self energies show a strong
isovector part by including a $\delta$-meson which is usually neglected in relativistic meson
exchange models.

This investigation is in some sense an extension of earlier non-relativistic work in Refs.~\cite{BSV.08}. We want to point out, however, that in the present work we were forced to leave, in part, the strategy of Ref.~\cite{BSV.08}. In that paper and in earlier work of Fayans~\cite{Fay.98} the strict Kohn-Sham strategy was followed as in Coulombic systems. Namely the bulk part was exclusively determined from previous microscopic Brueckner calculations~\cite{BMS.04} and thus fixed once and for all. Then a phenomenological finite range contribution and a spin-orbit term was added to the functional to account for properties of finite nuclei in adjusting four parameters. Since in the relativistic case the spin-orbit is fixed already from the nuclear matter calculations~\cite{Wal.74}, if all the parameters of the relativistic approach which survive in the infinite matter limit were adjusted to microscopic infinite matter results, one would essentially remain with only one adjustable parameter, i.e. the $\sigma$-mass, which can serve for the adjustment to properties of finite nuclei. The $\omega$- and the $\rho$-mass cannot be fixed independently from present data and therefore they are kept at their experimental values in the vacuum. It turns, however, out that only one parameter is not enough to reach the required accuracy for nuclear masses and radii. We, therefore, had to adjust parameters simultaneously to microscopic bulk properties and experimental finite nuclei data, thus departing from the strict Kohn-Sham strategy. Nevertheless, we ensured that in our model the density dependence of the meson exchange couplings is completely determined by the microscopic calculations of infinite matter and only four remaining parameters are adjusted to experimental data in finite nuclei.

On the other side we have to keep in mind that the nuclear many-body problem is much more complicated
than Coulombic systems. It is based on QCD, a relativistic theory, where spin-degrees of freedom play
an essential role, not only in the spin-orbit term. Relativistic models provide a consistent treatment
of the spin degrees of freedom and velocity dependent terms, they include the complicated interplay
between the large Lorentz scalar and vector self-energies induced on the QCD level by the in medium
changes of the scalar and vector quark condensates~\cite{CFG.92}. In particular they include nuclear
magnetism, i.e. the nuclear currents induced by the spatial parts of the vector self energies
or the time-odd components of the nuclear density density functional. The Kohn-Sham
strategy applied in Ref.~\cite{BSV.08} has, at present, no possibility to derive this part
of the functional.

As a result of our investigations we have derived a functional DD-ME$\delta$ with similar properties as the very successful functional DD-ME2. In contrast to that model, DD-ME$\delta$ is based to a large extent on microscopic calculations. Only four parameters had to be adjusted to finite nuclei.  It turns out that the inclusion of the $\delta$-meson does not improve the accuracy of the properties of finite nuclei such as masses and radii. Therefore the corresponding vertex and its density dependence is completely determined by nuclear matter data such as the isovector part of the effective Dirac mass. It is nevertheless much more physical and notably the
mass splitting of neutrons and protons is now correctly incorporated. It has, moreover, an influence on the behavior of the equation of state at higher densities and we find in this region a much better agrement with experimental data derived from heavy ion reactions~\cite{DLL.02} than the earlier parameter set DD-ME2 not including the $\delta$-meson. We therefore can hope that the new parameter set derived in this investigation is more reliable for applications of relativistic density functional theory to neutron stars.



\begin{acknowledgments}
We thank M. Baldo, G. A. Lalazissis, L. M. Robledo, B. K. Sharma, and D. Vretenar 
for helpful discussions
and C. Fuchs for the release of detailed results of his relativistic
Brueckner calculations in Ref.~\cite{DFF.07}.
X.R., X.V., and M.C. acknowledge partial support by the Spanish
Consolider-Ingenio 2010 Programme CPAN CSD2007-00042 and by Grants
No.\ FIS2008-01661 from MICINN (Spain) and FEDER and No.\ 2009SGR-1289
from Generalitat de Catalunya (Spain). We also acknowledge partial support
from the IN2P3-CAICYT collaboration (ACI10-00592) and from the
DFG Cluster of Excellence \textquotedblleft Origin and Structure of the
Universe\textquotedblright\ (www.universe-cluster.de).
\end{acknowledgments}


\newpage
\begin{table}[h!]
\begin{center}
\begin{tabular}{lrrrrrcclrrrrrcclrrrrr}
\hline\hline
   &     &          &          &       &       & & &    &     &          &          &       &       & & &    &     &          &          &       &      \\
El.&$N$&$B\quad$&$B^{\rm exp.}$&$r_{\rm c}$&$r^{\rm exp.}_{\rm c}$ &
& & El.&$N$&$B\quad$&$B^{\rm exp.}$&$r_{\rm c}$&$r^{\rm exp.}_{\rm
c}$ & & &
El.&$N$&$B\quad$&$B^{\rm exp.}$&$r_{\rm c}$&$r^{\rm exp.}_{\rm c}$ \\
   &     &[MeV]     & [MeV]    & [fm]  & [fm]  & & &    &     & [MeV]    & [MeV]    & [fm]  & [fm]  & & &    &     & [MeV]    & [MeV]    & [fm]  & [fm] \\
   &     &          &          &       &       & & &    &     &          &          &       &       & & &    &     &          &          &       &      \\
\hline
   &     &          &          &       &       & & &    &     &          &          &       &       & & &    &     &          &          &       &      \\
Ne &   6 &   99.266 &   97.321 & 3.270 & --    & & &    &  50 &  783.457 &  783.892 & 4.253 & 4.269 & & & Dy &  82 & 1209.677 & 1210.780 & 5.011 & 5.054\\
   &   8 &  136.762 &  132.143 & 3.004 & 2.972 & & &    &  52 &  798.416 &  799.721 & 4.275 & 4.306 & & & Er &  82 & 1213.553 & 1215.331 & 5.046 & 5.039\\
   &  16 &  201.409 &  201.601 & 2.934 & 2.927 & & & Mo &  44 &  722.829 &  725.831 & 4.282 & --    & & & Yb &  82 & 1215.892 & 1218.382 & 5.079 & 5.030\\
   &  18 &  210.055 &  206.929 & 2.960 & 2.963 & & &    &  46 &  748.226 &  750.117 & 4.289 & --    & & & Pt &  92 & 1326.638 & 1327.406 & 5.285 & --   \\
   &  20 &  216.614 &  211.214 & 2.987 & --    & & &    &  48 &  772.492 &  773.727 & 4.296 & --    & & &    &  94 & 1347.112 & 1348.344 & 5.297 & --   \\
Mg &   8 &  139.397 &  134.468 & 3.262 & --    & & &    &  50 &  795.461 &  796.508 & 4.302 & 4.316 & & & Hg &  92 & 1327.960 & 1326.766 & 5.310 & --   \\
   &  20 &  252.470 &  249.849 & 3.095 & --    & & &    &  52 &  812.048 &  814.255 & 4.325 & 4.352 & & &    &  94 & 1349.499 & 1348.469 & 5.322 & --   \\
Si &  20 &  284.656 &  283.429 & 3.160 & --    & & & Ru &  50 &  805.152 &  806.848 & 4.348 & --    & & &    &  96 & 1370.090 & 1369.743 & 5.334 & --   \\
   &  22 &  294.829 &  292.097 & 3.186 & --    & & &    &  52 &  823.450 &  826.495 & 4.371 & 4.393 & & &    & 124 & 1604.479 & 1608.651 & 5.487 & 5.474\\
 S &  20 &  308.727 &  308.714 & 3.272 & 3.298 & & & Pd &  50 &  812.826 &  815.088 & 4.392 & --    & & &    & 126 & 1617.160 & 1621.049 & 5.494 & 5.485\\
Ar &  20 &  329.008 &  327.342 & 3.357 & 3.402 & & &    &  52 &  832.905 &  836.301 & 4.414 & --    & & & Pb &  96 & 1371.354 & 1368.974 & 5.354 & --   \\
   &  22 &  345.907 &  343.810 & 3.372 & 3.427 & & & Cd &  50 &  818.559 &  821.067 & 4.432 & --    & & &    &  98 & 1392.218 & 1390.624 & 5.365 & --   \\
Ca &  16 &  281.114 &  281.360 & 3.426 & --    & & &    &  52 &  840.493 &  843.829 & 4.453 & --    & & &    & 100 & 1412.482 & 1411.654 & 5.376 & --   \\
   &  18 &  314.695 &  313.122 & 3.423 & --    & & & Sn &  50 &  822.373 &  824.794 & 4.468 & --    & & &    & 102 & 1432.207 & 1432.015 & 5.386 & --   \\
   &  20 &  345.755 &  342.052 & 3.425 & 3.476 & & &    &  52 &  846.232 &  849.086 & 4.490 & --    & & &    & 104 & 1451.439 & 1451.794 & 5.398 & --   \\
   &  22 &  366.416 &  361.895 & 3.436 & 3.506 & & &    &  54 &  869.413 &  871.891 & 4.510 & --    & & &    & 106 & 1470.212 & 1471.071 & 5.409 & --   \\
   &  24 &  385.355 &  380.959 & 3.449 & 3.515 & & &    &  56 &  892.038 &  893.867 & 4.528 & --    & & &    & 108 & 1488.548 & 1489.815 & 5.420 & 5.421\\
   &  26 &  402.728 &  398.769 & 3.464 & 3.493 & & &    &  58 &  913.812 &  914.626 & 4.546 & 4.561 & & &    & 110 & 1506.458 & 1508.096 & 5.431 & 5.429\\
   &  28 &  417.267 &  415.990 & 3.479 & 3.474 & & &    &  60 &  934.038 &  934.571 & 4.561 & 4.581 & & &    & 112 & 1523.947 & 1525.891 & 5.442 & 5.436\\
   &  30 &  428.933 &  427.490 & 3.501 & 3.514 & & &    &  62 &  953.239 &  953.531 & 4.576 & 4.594 & & &    & 114 & 1541.017 & 1543.186 & 5.452 & 5.442\\
   &  32 &  438.705 &  436.571 & 3.525 & --    & & &    &  64 &  971.484 &  971.574 & 4.590 & 4.610 & & &    & 116 & 1557.666 & 1560.019 & 5.463 & 5.450\\
Ti &  18 &  316.303 &  314.491 & 3.572 & --    & & &    &  66 &  988.808 &  988.684 & 4.604 & 4.627 & & &    & 118 & 1573.883 & 1576.354 & 5.473 & 5.459\\
   &  20 &  350.990 &  346.905 & 3.540 & --    & & &    &  68 & 1005.287 & 1004.954 & 4.618 & 4.641 & & &    & 120 & 1589.641 & 1592.187 & 5.483 & 5.469\\
   &  22 &  374.927 &  375.475 & 3.535 & --    & & &    &  70 & 1021.023 & 1020.546 & 4.633 & 4.654 & & &    & 122 & 1604.897 & 1607.506 & 5.492 & 5.479\\
   &  26 &  418.185 &  418.699 & 3.549 & 3.591 & & &    &  72 & 1036.100 & 1035.529 & 4.647 & 4.666 & & &    & 124 & 1619.568 & 1622.324 & 5.501 & 5.490\\
   &  28 &  436.198 &  437.781 & 3.559 & 3.570 & & &    &  74 & 1050.567 & 1049.963 & 4.661 & 4.676 & & &    & 126 & 1633.472 & 1636.430 & 5.509 & 5.501\\
   &  30 &  450.512 &  451.962 & 3.582 & --    & & &    &  76 & 1064.429 & 1063.889 & 4.676 & --    & & &    & 128 & 1642.984 & 1645.552 & 5.529 & 5.523\\
   &  32 &  463.004 &  464.234 & 3.606 & --    & & &    &  78 & 1077.635 & 1077.346 & 4.690 & --    & & &    & 130 & 1652.223 & 1654.514 & 5.548 & 5.545\\
Cr &  22 &  380.810 &  381.978 & 3.625 & --    & & &    &  80 & 1090.071 & 1090.293 & 4.703 & --    & & &    & 132 & 1661.295 & 1663.291 & 5.565 & 5.565\\
   &  28 &  452.793 &  456.349 & 3.627 & 3.642 & & &    &  82 & 1101.452 & 1102.851 & 4.715 & --    & & & Po & 120 & 1597.316 & 1599.165 & 5.517 & 5.503\\
Fe &  28 &  466.874 &  471.763 & 3.687 & 3.693 & & &    &  84 & 1108.261 & 1109.235 & 4.736 & --    & & &    & 122 & 1613.484 & 1615.156 & 5.526 & 5.512\\
   &  38 &  552.272 &  550.994 & 3.801 & --    & & & Te &  74 & 1065.370 & 1066.368 & 4.708 & 4.727 & & &    & 124 & 1629.025 & 1630.586 & 5.535 & 5.522\\
   &  40 &  564.993 &  561.939 & 3.820 & --    & & &    &  76 & 1080.802 & 1081.439 & 4.722 & 4.735 & & &    & 126 & 1643.753 & 1645.212 & 5.542 & 5.534\\
   &  42 &  575.189 &  571.637 & 3.840 & --    & & &    &  78 & 1095.581 & 1095.941 & 4.735 & 4.743 & & &    & 128 & 1654.405 & 1655.772 & 5.562 & --   \\
Ni &  26 &  448.501 &  453.156 & 3.735 & --    & & &    &  80 & 1109.632 & 1109.914 & 4.747 & --    & & &    & 130 & 1664.760 & 1666.015 & 5.581 & --   \\
   &  28 &  477.871 &  483.992 & 3.735 & --    & & &    &  82 & 1122.766 & 1123.434 & 4.759 & --    & & &    & 132 & 1674.904 & 1675.904 & 5.599 & --   \\
   &  30 &  500.345 &  506.458 & 3.763 & 3.775 & & &    &  84 & 1130.563 & 1131.442 & 4.781 & --    & & & Rn & 122 & 1620.888 & 1621.200 & 5.559 & 5.535\\
   &  38 &  577.084 &  576.808 & 3.848 & --    & & & Xe &  80 & 1127.656 & 1127.434 & 4.789 & 4.792 & & &    & 124 & 1637.270 & 1637.293 & 5.567 & 5.544\\
   &  40 &  592.287 &  590.408 & 3.866 & --    & & &    &  82 & 1142.487 & 1141.877 & 4.799 & 4.799 & & &    & 126 & 1652.787 & 1652.497 & 5.574 & 5.554\\
   &  42 &  604.799 &  602.236 & 3.886 & --    & & &    &  84 & 1151.416 & 1151.746 & 4.822 & 4.836 & & &    & 128 & 1664.614 & 1664.300 & 5.594 & --   \\
   &  44 &  615.703 &  613.169 & 3.906 & --    & & & Ba &  80 & 1144.124 & 1142.775 & 4.828 & 4.833 & & &    & 130 & 1676.101 & 1675.867 & 5.613 & --   \\
Zn &  28 &  480.525 &  486.964 & 3.840 & --    & & &    &  82 & 1160.594 & 1158.292 & 4.838 & 4.838 & & & Ra & 122 & 1627.084 & 1625.669 & 5.590 & 5.554\\
   &  40 &  610.734 &  611.086 & 3.938 & 3.985 & & &    &  84 & 1170.762 & 1169.444 & 4.862 & 4.870 & & &    & 124 & 1644.283 & 1642.464 & 5.598 & 5.562\\
   &  42 &  625.597 &  625.796 & 3.955 & --    & & & Ce &  80 & 1158.826 & 1156.034 & 4.865 & 4.873 & & &    & 126 & 1660.562 & 1658.315 & 5.605 & 5.571\\
Ge &  50 &  700.989 &  702.437 & 4.080 & --    & & &    &  82 & 1176.896 & 1172.692 & 4.875 & 4.877 & & &    & 128 & 1673.586 & 1671.267 & 5.625 & --   \\
Se &  50 &  727.463 &  727.343 & 4.126 & --    & & &    &  84 & 1188.434 & 1185.289 & 4.899 & 4.907 & & &    & 130 & 1686.209 & 1684.050 & 5.644 & --   \\
   &  52 &  737.305 &  738.074 & 4.153 & --    & & &    &  86 & 1199.632 & 1197.330 & 4.921 & 4.931 & & & Th & 122 & 1632.040 & 1628.617 & 5.621 & --   \\
Kr &  50 &  749.164 &  749.234 & 4.169 & 4.184 & & & Nd &  80 & 1168.468 & 1167.295 & 4.901 & 4.910 & & &    & 124 & 1650.041 & 1646.139 & 5.628 & --   \\
   &  52 &  760.633 &  761.804 & 4.195 & 4.217 & & &    &  82 & 1187.830 & 1185.141 & 4.909 & 4.912 & & &    & 126 & 1667.060 & 1662.689 & 5.634 & --   \\
Sr &  48 &  749.276 &  748.928 & 4.201 & 4.226 & & &    &  84 & 1200.491 & 1199.082 & 4.932 & 4.941 & & &    & 128 & 1681.298 & 1676.762 & 5.655 & --   \\
   &  50 &  767.989 &  768.468 & 4.209 & 4.220 & & & Sm &  80 & 1176.227 & 1176.614 & 4.935 & 4.944 & & &    & 130 & 1695.061 & 1690.610 & 5.673 & --   \\
   &  52 &  781.200 &  782.631 & 4.233 & 4.261 & & &    &  82 & 1196.890 & 1195.736 & 4.942 & 4.944 & & &    & 126 & 1672.236 & 1665.648 & 5.664 & --   \\
Zr &  46 &  740.699 &  740.644 & 4.237 & --    & & &    &  84 & 1210.720 & 1210.909 & 4.965 & 4.975 & & &  U & 132 & 1717.151 & 1710.285 & 5.719 & --   \\
   &  48 &  762.784 &  762.605 & 4.246 & 4.281 & & & Gd &  82 & 1204.158 & 1204.435 & 4.977 & 4.976 & & &    &     &          &          &       &      \\
   &     &          &          &       &       & & &    &     &          &          &       &       & & &    &     &          &          &       &      \\
\hline\hline
\end{tabular}
\caption{DD-ME$\delta$ results for the binding energies and charge radii used in the fit}%
\label{tab5}
\end{center}
\end{table}

\end{document}